\documentclass[preprint,12pt]{elsarticle}

\usepackage{epstopdf}
\usepackage{amsmath}
\usepackage{amsfonts}
\usepackage[dvipsnames]{xcolor}
\usepackage{color}
\usepackage{psfrag}

\usepackage{subfigure}
\usepackage{float}

\newtheorem{theorem}{Theorem}
\newtheorem{remark}{Remark}

\newtheorem{assumption}{Assumption}

\newenvironment{proof}{{\bf Proof:}}{\hfill $\square$}

\newcommand\redsout{\bgroup\markoverwith
	{\textcolor{red}{\rule[0.55ex]{2pt}{1.5pt}}}\ULon}

\usepackage{natbib}

\begin{document}

\begin{frontmatter}

\title{Robust stability analysis of an energy-efficient control in a Networked Control System with application to Unmanned Ground Vehicles\tnoteref{label1}}

\tnotetext[label1]{a) This manuscript is the
authors' original work and has not been published nor has it been
submitted simultaneously elsewhere. b) All authors have checked the manuscript and have agreed to the submission. The corresponding author is A. Gonz\'alez. A. Gonz\'alez, A. Cuenca, J. Salt are with the Instituto de Autom\'{a}tica e Inform\'{a}tica Industrial (AI2), Universitat Polit\`{e}cnica de Val\`{e}ncia (UPV), Spain. (e-mail: angonsor@upv.es,acuenca@isa.upv.es,julian@isa.upv.es). J. Jacobs is with the Department of Mechanical Engineering, Eindhoven University of Technology, Eindhoven (The Netherlands); jelle\_jacobs@hotmail.nl
}

\author{Antonio Gonz\'alez, \'Angel Cuenca, Juli\'an Salt, Jelle Jacobs}

\begin{abstract}
In this paper, the robust stability and disturbance rejection performance analysis of an energy-efficient control is addressed in the framework of Networked Control System (NCS). The control scheme under study integrates periodic event-triggered control, packet-based control, time-varying Kalman filter, dual-rate control and prediction techniques, whose design is aimed at reducing energy consumption and bandwidth usage. The robust stability against time-varying model uncertainties is analyzed by means of a sufficient condition based on Linear Matrix Inequalities (LMI). Finally, the effectiveness of the proposed approach is experimentally validated in a tracking control for an Unmanned Ground Vehicle (UGV), which is a battery-constrained mobile device with limited computation capacities.

\end{abstract}
\end{frontmatter}

\section{Introduction}
The research of control synthesis methods for increasing the energy-efficiency of the control system has been a subject of research for the last years \cite{xia2010energy,xia2016industrial}. An appropriate design for efficiently controlling energy-consuming systems is a critical aspect in many control engineering applications, such as heating \cite{vsiroky2011experimental}, chemical processes \cite{simkoff2020process}, electric vehicles \cite{han2018energy}, solar energy systems \cite{zou2016energy}, where many of them are implemented in Networked Control Systems (NCS) \cite{andrade2016evaluation,park2019robust}. 

NCS are control systems in which the plant to be controlled and the rest of control devices are spatially distributed, and hence, the communication among them occurs through a shared and band-limited digital communication network \cite{zhang2019networked}. Taking into account the limited communication bandwidth, the use of event-triggered control (ETC) for data transmission are advantageous with respect to time-triggered protocols \cite{heemels2012periodic,li2020event}, since data packets are only transmitted when certain event conditions are satisfied. This feature has motivated the implementation of ETC approaches into NCS designs under different scenarios, such as the presence of denial-of-service attacks \cite{tripathy2019robust,guo2020event}, aperiodic sampling schemes \cite{tripathy2020robust} and multiagent systems \cite{luo2021event,zhang2021leader}, among others. In particular, 
wireless communication networks are sometimes preferred, especially in a context where it is quite expensive and difficult to install wired connections. However, network devices are usually powered by means of batteries with limited capacity, so wireless data transmission may become very expensive in terms of energy consumption \cite{raghunathan2002energy}. Hence, achieving a reliable wireless communication through a suitable NCS design aimed at reducing energy waste is crucial in order to maximize network lifetime \cite{alcaina2019energy}. In this kind of setups, the benefits of ETC has been discussed not only to save bandwidth resources but also to improve energy efficiency \cite{cuenca2018non,gonzalez2019event,ding2020event}.



Another useful solution that contributes to reduce bandwidth and energy consumption in wireless NCS is to employ different rates for sensor and actuator. In this method, known in the literature as dual-rate control \cite{li2002analysis}, measurement data are acquired at slow rate, and control actions are injected at fast rate in order to improve closed-loop performance to some extent (see, e.g., \cite{salt2005mbm}). This kind of control solution, combined with packet-based control strategies (see, e.g., \cite{cuenca2018packet}), enables to only send data through the network at the slow rate.

Apart from energy and bandwidth constraints, other negative effects inherent to NCS are time-varying communications delays, packet dropouts and packet disorder. Such phenomena have been widely investigated in the literature. Time-varying delays in the control system has been tackled under different approaches: state-feedback control \citep{cloosterman2009stability}, state estimators \cite{penarrocha2012state}, multi-rate control  \cite{sala2009retunable} and predictor-feedback control approaches \cite{gonzalez2021weighted}. Further extensions were adapted to deal with time-varying delays together with packet dropouts by gain-scheduling predictor-feedback approaches \cite{gonzalez2019gain,gonzalez2019event}, and active disturbance rejection by integrating an extended state observer \cite{hao2019output}. Packet dropouts have also been faced using predictive control \cite{li2014network}, gain scheduling \cite{dolz2016networked}, and predictor-observer methods \cite{cuenca2018non}. In the latter case, the underlying idea is to reconstruct the actuator and sensor signals by combining predictor approaches with a Luenberger observer. Other alternative observer-based methods resort to proportional multi-integral observers for signal reconstruction in the presence of actuator and sensor faults \cite{kuhne2018fault}. Lastly, packet disordering has also been addressed by introducing different packet reordering mechanisms \cite{liu2015new,liu2017networked}, and by means of dual-rate control \cite{sala2009retunable,cuenca2018periodic,cuenca2018packet}.

Very recently, an energy-efficient control strategy was proposed in \cite{alcaina2019energy} applied to an output-feedback tracking control of a UGV in a wireless sensor network. The proposed control scheme combines event-triggered protocols and dual-rate control with the objective of reducing energy and bandwidth consumption. Moreover, a predictor-based observer was integrated with a Time-Varying Dual-Rate Kalman Filter (TVDRKF) in order to deal with packet dropouts and time-varying delays. This work provides simulation results performed by means of a Truetime application \cite{cervin2003does}, showing the achieved improvements in terms of energy efficiency and reduction of bandwidth usage. Nevertheless, to the best of the authors' knowledge, two aspects have been not addressed in previous related works: (i) a formal analysis of the closed-loop control performance in terms of robust stability against model mismatches and disturbance rejection as a function of the event-triggered thresholds, and (ii) an experimental validation of the proposed NCS design. Indeed, demonstrators of ETC implemented in NCS are rare, where some exceptions can be found in \cite{gonzalez2019event,dolk2017event,khashooei2017suboptimal,shah2020event,cuenca2018periodic}.

In this paper, the stability analysis is carried out via Lyapunov approaches and robust control theory. As a result, a sufficient condition is obtained to ascertain the robust stability of the closed-loop control system with guaranteed disturbance rejection index in terms of Linear Matrix Inequalities (LMI) \cite{boyd1994linear}, which can be efficiently solved using available semidefinite programming tools. Moreover, experimental data is also provided to validate the effectiveness of the control design in a prototype consisting in a UGV, which is a two-wheel Lego Mindstorms EV3 robot equipped with a wifi-dongle to send and receive data-packages through the wireless network.

The paper is structured as follows: Section \ref{sec:intro} describes the problem statement and gives some preliminaries. Section \ref{sec:control} introduces the proposed control strategy. Section \ref{sec:analysis} addresses the robust stability analysis of the closed-loop system. Section \ref{sec:UGV} presents the application of the control solution to a UGV, studying the robust stability of the control system, and providing simulation results and their experimental validation. Finally, some conclusions are gathered in Section \ref{sec:conc}.
 
\section{Problem statement and preliminaries}\label{sec:intro}
Let $G_p(s)$ be the transfer function of the plant system to be controlled. A discrete-time state-space representation of the discretized system $G_p(s)$ with zero-order hold (ZOH) at sampling period $T$ considering model uncertainties yields:
\begin{align}\label{eq:model}
&x_{p,k+1}^{T} = \left(A_p + \Delta_{A,k}\right) x_{p,k}^{T} + \left(B_p + \Delta_{B,k}\right) \left(u_{k}^{T} + d_{k}^{T}\right), \\ \nonumber
&y_k^{T} = C_p x_{p,k}^{T} + v_{k}^{T}
\end{align}
where $x_{p,k}^{T} \in \mathcal{R}^{n_p}$ is the state vector containing $n_p$ state variables of the plant system, $u_k^{T} \in \mathcal{R}^{m}$ is the control input  with $m$ input signals, $d_k^{T} \in \mathcal{R}^{m}$ is a matched disturbance input (assumed to be unmeasurable), $y_k^{T} \in \mathcal{R}^{q}$ represents the output system with $q$ output signals, $v_{k}^{T} \in \mathcal{R}^{q}$ is the measurement noise, $A_p, B_p, C_p$ are the state-space matrices of appropriate dimensions, and $\Delta_{A,k}, \Delta_{B,k}$ are time-varying model uncertainties described below in \eqref{eq:uncertt}.

In this paper, dual-rate control is used in the controller scheme with two different periods: $T$ as the actuation period, and $NT$ as the sensing period, being $N \in \mathbb{N^+}$ the multiplicity between the two periods of the dual-rate control scheme. Thereafter, let us respectively introduce the notation $(.)_k^{T}$ and $(.)_k^{NT}$ to denote a $T$-period and an $NT$-period signal or variable, where $k \in \mathbb{N}$ are iterations at the corresponding period.

\begin{figure}\centering
	\includegraphics[width=14cm, height=8cm]{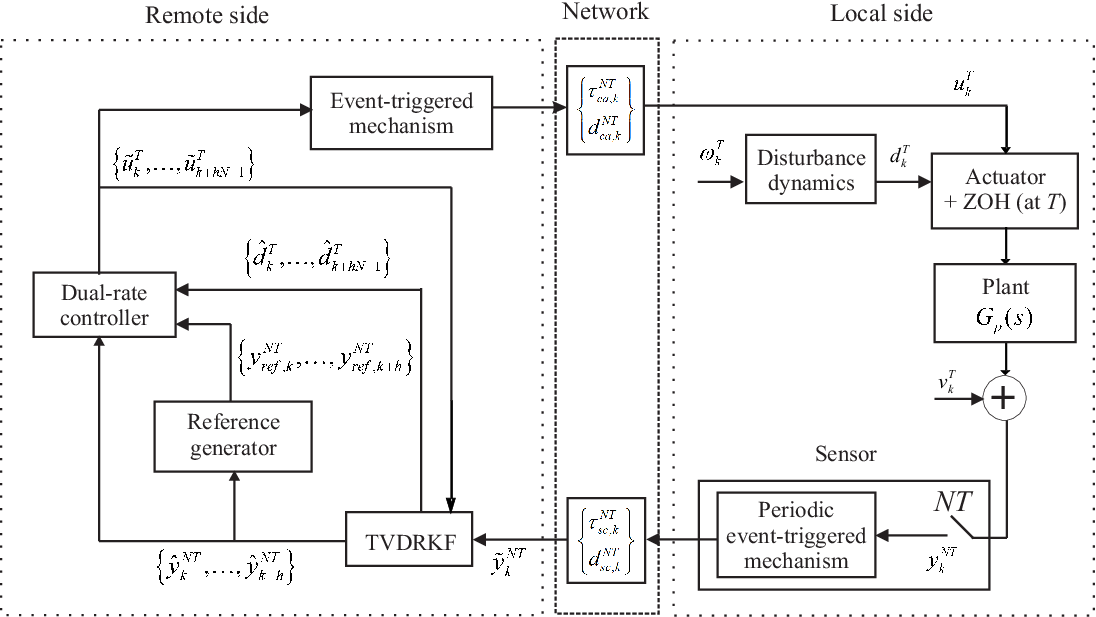}
	\caption{Proposed control scheme for the NCS}\label{fig:ControlScheme}
\end{figure}

Consider the NCS described in Fig. \ref{fig:ControlScheme}, where two main components can be distinguished: local side and remote side. The local side includes the sensor, actuator, and the plant system $G_p(s)$ to be controlled. As computational resource constraints are assumed at the local side, the digital control algorithm is located in a computationally powerful device at the remote side. Remote and local sides are connected by a shared communication network, where communication delays and packet dropouts may occur. For a packet sampled at instant $kNT$, being effectively sent (i.e., the trigger conditions hold) and not lost, the round-trip time delay is defined as $\tau_{k}^{NT} = \tau_{ca,k}^{NT} + \tau_{sc,k}^{NT} + \tau_{c,k}$, where $\tau_{ca,k}^{NT}$ and $\tau_{sc,k}^{NT}$ are the communication delay induced by the controller-to-actuator and sensor-to-controller channels, respectively, and $\tau_{c,k}$ is a computation time delay. Packet dropouts are contemplated as a Bernoulli process, whose probability of dropout is respectively given by $p_{sc}$ and $p_{ca}$:
\begin{equation}\label{probability}
\begin{split}
p_{sc}=\mathrm{Pr}[d^{NT}_{sc,k}=0] \in [0,1)  \\ p_{ca}=\mathrm{Pr}[d^{NT}_{ca,k}=0] \in [0,1)
\end{split}
\end{equation}

The structure of the output predictor-feedback control $u_k^T$ is designed with the objective of reducing energy and bandwidth consumption. For this purpose, a periodic event-triggered control is combined with packet-based control, time-varying dual-rate Kalman filter, prediction techniques, and dual-rate control. A detailed description is later presented in Section \ref{sec:control}.

The objective of this paper is two-fold: (i) to provide sufficient conditions to ascertain the robust stability of the designed NCS with guaranteed disturbance rejection index, and (ii) validate the control strategy in an experimental setup consisting in a tracking control of a UGV.

The following assumptions and preliminary results are given for the stability analysis and control synthesis:
\begin{assumption}\label{as0}
	The disturbance $d_k^T \in \mathcal{R}^m$ can be modeled by the following exogenous system \cite{hao2019output}:
	\begin{align}\label{eq:dmod}
	&x_{d,k+1}^T = A_d  x_{d,k}^T + B_d \omega_k^T, \\ \nonumber
	&d_k^T = C_d x_{d,k}^T
	\end{align}
	where $A_d \in \mathcal{R}^{r \times r}$, $B_d \in \mathcal{R}^{r \times m}$, $C_d \in \mathcal{R}^{m \times r}$ are known matrices, the initial conditions $x_{d,0}$ is assumed to be unknown, and $\omega_k \in \mathcal{R}^q$ represents an unknown disturbance component. This assumption allows modeling some specific type of disturbances typically encountered in many industrial applications, such as harmonics or unknown load disturbances \cite{guo2005disturbance}.
\end{assumption}

\begin{assumption}\label{as1}
	Time-varying model uncertainties are assumed to be described as a norm-bounded form \cite{han2001robust}:
	\begin{align}\label{eq:uncertt}
	\left(\Delta_{A,k},  \Delta_{B,k} \right) = \delta_{\Delta} E_p \Delta_k \left(H_{p,A},  H_{p,B} \right)
	\end{align}
	where $\delta_{\Delta} \geq 0$ is a scalar that determines the size of uncertainties, $\Delta_k \in \mathcal{R}^{l_1 \times l_2}$ represents any unknown time-varying matrix satisfying $\Delta_k^{'} \Delta_k \leq I, \forall k \geq 0$, where hereinafter the notation $(\cdot)^\mathrm{'}$ denotes the transpose function, and $E_p \in \mathcal{R}^{n_p \times l_1}$, $H_{p,A} \in \mathcal{R}^{l_2 \times n_p}$, $H_{p,B} \in \mathcal{R}^{l_2 \times m}$ are time-constant matrices that define the structure of such uncertainties. The norm-bounded form \eqref{eq:uncertt} can be understood as an ellipsoid around the nominal state-space matrices whose shape is defined through matrices $E_p, H_{p,A}, H_{p,B}$, and whose size is determined by the scalar $\delta_{\Delta}$. This model is widely used in the literature to describe uncertain systems with time-varying model mismatches, and may include parametric uncertainties, unmodeled dynamics, small variations in the sampling period \cite{suh2008stability} and in general all possible source of bounded uncertainties. Note that \eqref{eq:uncertt} can also describe unmatched disturbances by choosing, for instance, $E_p = diag\left(E_{p1}, E_{p2}\right)$ and $\Delta_k= \left(\Delta_{1,k}, \Delta_{2,k}\right)$ with uncorrelated time-varying matrices $\Delta_{1,k}$, $\Delta_{2,k}$. In this case, $\Delta_{A,k}$ and $\Delta_{B,k}$ are not necessarily in the same range space as the nominal input matrix $B_p$ \cite{tripathy2020robust}.
\end{assumption}

\begin{remark}\label{as2}
	The sampling period at slow rate is chosen to be larger than the largest round-trip time delay $\tau_{max} = max\left(\tau_{k}^{NT}\right), \forall k \geq 0$ in order to avoid packet disorder, that is to say, $NT > \tau_{max}, \forall k \geq 0$. The largest round-trip delay can be available by assuming prior knowledge of a statistical distribution for the network-induced delay \cite{bolot1993end}.
\end{remark}

\begin{remark}\label{remark:aummm}
	System $x_{k+1}^{NT} = A x_k^{NT} + B u_k^{NT}, \ y_k^{NT} = C x_k^{NT}$ with $A \in \mathcal{R}^n$, $B \in \mathcal{R}^{n \times m}$ and $C \in \mathcal{R}^{q \times n}$ can equivalently be written at fast period $T$ as $x_{k+N}^{T} = A x_k^{T} + B u_k^{T}, \ y_k^T = C x_k^{T}$, and in augmented form at fast period as:
	\begin{align} \nonumber
	&\bar{x}_{k+1}^T = \bar{A} \bar{x}_k^T + \bar{B} u_k^T, \\ 
	&y_k^T = \bar{C} \bar{x}_k^T
	\end{align}
	where $\bar{x}_k^T = \begin{bmatrix} \left(x_{k-1}^T\right)^{'} & \  \left(x_{k-2}^T\right)^{'} & \ \cdots &  \left(x_{k-N}^T\right)^{'} \end{bmatrix}^{'}$ and
	\begin{align}
	\bar{A} = \begin{bmatrix}0 & A \\ I_{(N-1)n} & 0  \end{bmatrix}, \ \bar{B} = \begin{bmatrix}B \\ 0_{(N-1)n \times m} \end{bmatrix}, \ \bar{C} = \begin{bmatrix}0_{q \times (N-1)n} & C \end{bmatrix}
	\end{align}
\end{remark}

\section{Proposed control scheme}\label{sec:control}
This section presents the proposed control system (see Figure \ref{fig:ControlScheme}). Both in remote and local sides, event-triggered protocols are designed in order to decide when the data must be sent through the network. A detailed description of each component of the control scheme is provided in the next subsections.

\subsection{Event-triggered protocol for transmission of measurement data}\label{sec:evtmes}
The measurements of the output system $y_k^{T}$ are sent from sensors next to the plant system to the controller via network at slow period $NT$ under the following event-triggered mechanism:
\begin{align}\label{eq:y}
\tilde{y}_k^{NT} = \begin{cases} y_k^{NT} & \text{if \eqref{eq:cond2} is true} \\ \tilde{y}_{k-1}^{NT} & \text{otherwise}\end{cases}
\end{align}
The measurement $y_k^{NT}$ is therefore transmitted if the following event-triggered condition periodically evaluated at period $NT$ is satisfied:
\begin{align}\label{eq:cond2}
(y_k^{NT} - \tilde{y}_{k-1}^{NT})^{'} \Omega_y (y_k^{NT} - \tilde{y}_{k-1}^{NT}) > \sigma_y^2 \left(y_k^{NT}\right)^{'} \Omega_y y_k^{NT} + \delta_y,
\end{align}
where $\Omega_y \in \mathcal{R}^{p} > 0$, and $\sigma_y, \delta_y \in \mathcal{R} \geq 0$ are event-triggered parameters that define the threshold level to decide whether the measurement data packet must be sent or not to remote side.

\subsection{Event-triggered protocol for transmission of control actions}\label{sec:evtact}
Taking into account \eqref{eq:y} and \eqref{eq:cond2}, the outputs sampled at the slow rate $1/NT$ may arrive at the remote side at a slower rate $1/\bar{N}T$ in a non-uniform fashion. Let $\mu=\bar{N}/N$, where $\mu$ is the number of consecutive packet dropouts. Note that both $\mu$ and $\bar{N}$ will be time-varying due to the non-uniform nature of the time sampling pattern.
Consider the mixed event-triggered protocol to decide when the set of control actions $ \{\tilde{u}_k^T, \ \tilde{u}_{k+1}^T, \ ..., \ \tilde{u}_{k+hN-1}^T\}$ must be transmitted via network from the controller to the actuator, where $\tilde{u}_k^T$ is defined in \eqref{eq:fast}, and $h=max(\mu)$ is the maximum number of consecutive packet dropouts. The above set of control actions is transmitted if the following event-triggered condition is satisfied for the first control action:
\begin{align}\label{eq:cond1}
(\tilde{u}_k^T - u_{k-1}^T)^{'} \Omega_u (\tilde{u}_k^T - u_{k-1}^T) > \sigma_u^2 \left(\tilde{u}_k^{T}\right)^{'} \Omega_u \tilde{u}_k^T + \delta_u,
\end{align}
where \eqref{eq:cond1} is evaluated each time a new output measurement is received (i.e., when \eqref{eq:cond2} is satisfied), and $\Omega_u \in \mathcal{R}^{m} > 0$, and $\sigma_u, \delta_u \in \mathcal{R} \geq 0$ are event-triggered parameters that define the threshold level to decide whether the data packet containing the set of control actions must be sent or not to local side. Then, we have that:
\begin{align}\label{eq:u}
u_{k}^T = \begin{cases} \tilde{u}_{k}^T & \text{if \eqref{eq:cond1} is true} \\ u_{k-1}^T & \text{otherwise}\end{cases}
\end{align}

\subsection{Time-Varying Dual-Rate Kalman Filter (TVDRKF)}\label{sec:TVDRKF}

Let $x_k^T = \begin{bmatrix} \left(x_{p,k}^{T}\right)^{'} & \ \left(x_{d,k}^{T}\right)^{'}\end{bmatrix}'$. From \eqref{eq:model} and \eqref{eq:dmod}, an augmented state-space model can be obtained as:
\begin{align}\label{eq:plantd}
&x_{k+1}^T = A_k x_k^T + B_k u_k^T + B_{\omega} \omega_k^T, \\ \nonumber
&y_k^T = C x_k^T + v_k^T, \\ \nonumber
&d_k^T = \mathcal{C}_d x_k^T
\end{align}
where $A_k = A + \delta_{\Delta} E \Delta_k H_A$, $\ B_k = B + \delta_{\Delta} E \Delta_k H_{p,B}$, and
\begin{align}\label{eq:mattri}
&A = \begin{bmatrix} A_p & \ B_p C_d \\ 0 & \ A_d \end{bmatrix}, \quad B = \begin{bmatrix}B_p \\ 0 \end{bmatrix}, \quad B_{\omega} = \begin{bmatrix}0 \\ B_d \end{bmatrix}, \ C=\begin{bmatrix} C_p & \ 0 \end{bmatrix}, \\ \nonumber
& \mathcal{C}_d=\begin{bmatrix} 0 & \ C_d \end{bmatrix}, \quad E = \begin{bmatrix}E_p \\ 0 \end{bmatrix}, \quad H_A = \begin{bmatrix} H_{p,A} & \ 0 \end{bmatrix},
\end{align}

The best linear estimation for $x^T_k$ in the sense of mean square error is obtained by means of the conventional Kalman filter \cite{simon2006optimal} by considering $\omega_k^T$ and $v^T_k$ as zero-mean white noises for system \eqref{eq:plantd}. In case of having different rates in the control system, a multi-rate Kalman filter containing corrections and predictions is necessary \cite{zheng2016multi}. In the proposed control scheme, a dual-rate control and periodic ETC are implemented with the objective of minimizing bandwidth and energy consumption. Consequently, some measurements are not sent to the remote side, leading to a Time-Varying Dual Rate Kalman Filter (TVDRKF) \cite{trimpe2014event,cuenca2018non}.

In addition, from extended state observer \cite{zheng2016extended} and multi-rate Kalman filter \cite{zheng2016multi} methods, the proposed TVDRKF is able to estimate the plant measurement and disturbance by implementing the augmented system model given in \eqref{eq:plantd}.

Let $\hat{x}_{k|k}^T$ the estimation of the system state $x_k^T$ defined in \eqref{eq:plantd}. When the output measurement $\tilde{y}_k^{NT}$ (see \eqref{eq:y}) is available at the remote side, the TVDRKF can perform the correction (and filtering) stage:
\begin{align}\label{eq:kalm}
\hat{x}_{k|k}^{T} = \hat{x}_{k|k-\bar{N}}^{T} + K(\bar{N}) \left(\tilde{y}_k^{T} - C \hat{x}_{k|k-\bar{N}}^{T}\right)
\end{align}
where $\hat{x}_{k|k-\bar{N}}^{T}$ is the current prediction, which was obtained by iterating the prediction formula given below \eqref{eq:predy} $\bar{N}$ steps ahead from the estimation made at instant $(k-\bar{N})T$. When no measurement data is received, the above state estimation is updated at fast rate $T$ using  \eqref{eq:predy} for $l=1,2,..,\bar{N}_{max}$ with $\bar{N}_{max} = hN$.
\begin{align}\label{eq:predy}
\hat{x}_{k+l|k}^{T} = A^{l} \hat{x}_{k|k}^{T} + \sum_{c=0}^{l-1} A^{l-1-c} B \tilde{u}_{k+c}^T
\end{align}

The time-varying $K(\bar{N})$ can be calculated by multi-rate Kalman filter design techniques as
\begin{align}\label{K}
& K(\bar{N})=M_{k+1}C^\mathrm{'}[CM_{k+1}C^\mathrm{'}+V]^{-1}\\ \nonumber
& M_{k+1}=A^{\bar{N}}M_k(A^{\bar{N}})^\mathrm{'}+W_e-A^{\bar{N}}M_kC^\mathrm{'}[CM_kC^\mathrm{'}+V]^{-1}CM_k(A^{\bar{N}})^\mathrm{'}
\end{align}
where matrices $A, B, C$ are given in \eqref{eq:mattri}, $V=Cov(v_k^T)$ being $v_k^T$ the measurement noise, and
\begin{equation}\label{eq:lawe7}
\begin{split}
W_e&=Cov\left(\sum_{c=0}^{\bar{N}-1}A^{\bar{N}-1-c} B_{\omega}\omega_{k+c}^T \right) \\ 
&=\mathcal{E}\left\{\left(\sum_{c=0}^{\bar{N}-1}A^{\bar{N}-1-c} B_{\omega}\omega_{k+c}^T\right) \left(\sum_{c=0}^{\bar{N}-1}A^{\bar{N}-1-c} B_{\omega}\omega_{k+c}^T\right)^\mathrm{'}\right\} 
\end{split}
\end{equation}
where $\mathcal{E}\{\cdot\}$ denotes the expectation.
The process noise $\omega_k^T$ is assumed to be uncorrelated, i.e., $\mathcal{E}\left(\omega_{k+c}^T \omega_k^{T'}\right) = 0, \ \forall c \neq 0$. Hence, from \eqref{eq:lawe7} it can be deduced that
\begin{equation}
\begin{split}
W_e&= \displaystyle \sum_{c=0}^{\bar{N}-1} \left(A^{\bar{N}-1-c}B_{\omega} \right) W \left(A^{\bar{N}-1-c}B_{\omega}\right)^{'}
\end{split}
\end{equation}
where $W=Cov(\omega_k^T)$, being $\omega_k^T$ the process noise. 

Moreover, from \eqref{eq:predy}, the set of $hN$ estimated outputs and disturbances  can be calculated as follows:
\begin{equation}\label{estimates}
\begin{split}
\hat{y}^T_{k+l}&=C\hat{x}^T_{(k+l|k)}\\
\hat{d}^T_{k+l}&=\mathcal{C}_d \hat{x}^T_{(k+l|k)}
\end{split}
\end{equation}

\begin{figure}[H]
	\centering
	\includegraphics[width=10.0cm, height=7.0cm]{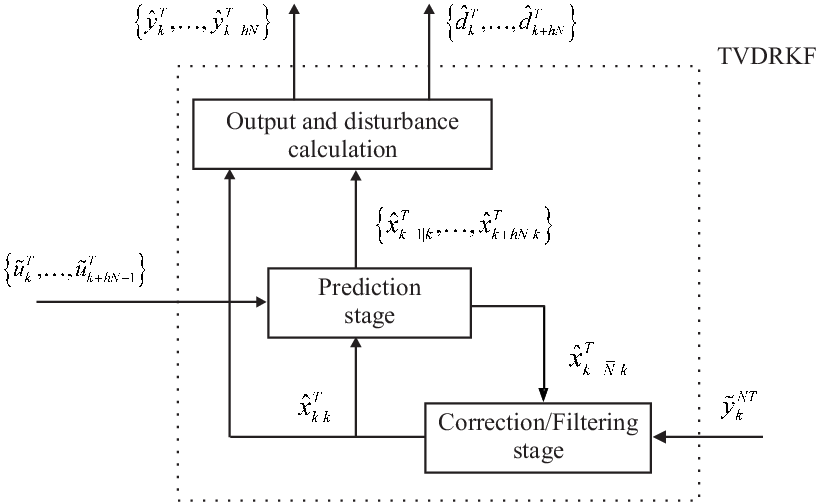}
	\caption{Structure of the Time-Varying Dual-Rate Kalman Filter}
	\label{fig:TVDRKF}
\end{figure}

Figure \ref{fig:TVDRKF} summarizes the structure of the TVDRKF.

\subsection{Dual-rate controller:}
\begin{figure}[H]
	\centering
	\includegraphics[width=1\linewidth]{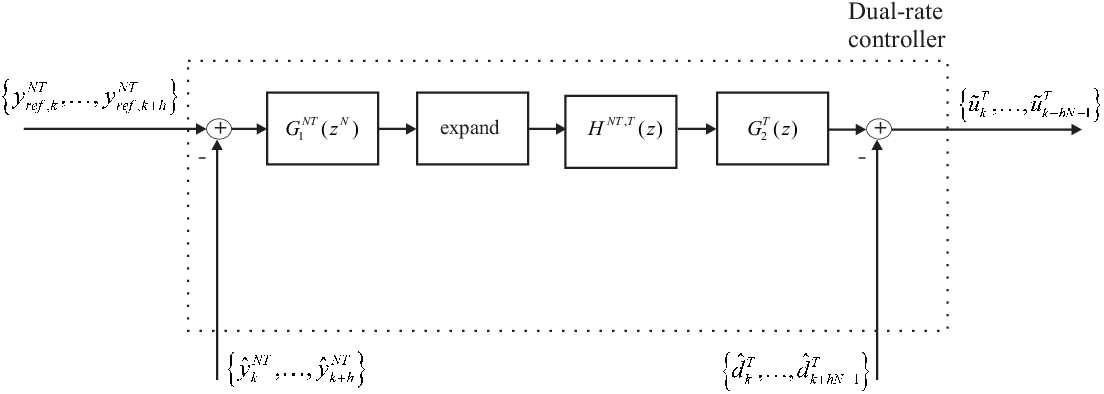}
	\caption{Dual-rate controller scheme}
	\label{fig:Dualrate}
\end{figure}
In this work we design a model-based dual-rate controller \cite{salt2005mbm}. Using $Z$-transform at period $T$, the dual-rate control structure is defined as follows (see Fig. \ref{fig:Dualrate}):
\begin{itemize}
	\item a slow-rate subcontroller
	$G_1^{NT}(z^N)=u_{1,k}^{NT}/e_k^{NT}$,
	\item a digital hold $H^{NT,T}(z)=u_{1,k}^T/[u_{1,k}^{NT}]^T$, and
	\item a fast-rate subcontroller $G_2^{T}(z)=\tilde{u}_{2,k}^T/u_{1,k}^T$.
\end{itemize}
where $z$ is the \textit{T}-unit operator. The input of $G_1^{NT}(z^N)$ is the tracking error $e_k^{NT} = y_{ref,k}^{NT} - \hat{y}_k^{NT}$, being $y_{ref,k}^{NT}$ the output tracking reference and $ \hat{y}_k^{NT}$ the estimation of the output system obtained by the TVDRKF explained in Section \ref{sec:TVDRKF}.

The dynamic dual-rate controller computes $N$ control actions at period $T$, which can be arranged in the augmented vector:
$$
\{\tilde{u}^T_{2,k},\tilde{u}^T_{2,k+1},\dots,\tilde{u}^T_{2,k+N-1}\}.
$$
The anti-disturbance approach is implemented by finally subtracting the disturbance estimations generated by the TVDRKF at period $T$, 
$$\{\hat{d}^T_k,\hat{d}^T_{k+1},\dots,\hat{d}^T_{k+N-1}\},
$$
which are obtained from \eqref{estimates}, leading to $\{\tilde{u}^T_{k}, \tilde{u}^T_{k+1},\dots, \tilde{u}^T_{k+N-1}\}$, where $\tilde{u}_{k+l}^T = \tilde{u}_{2,k+l}^T - \hat{d}^T_{k+l}$ with $l=0,1,...,N-1$. Following this operation mode for the next $h$ dynamic references and outputs, $\{\hat{y}^{NT}_{ref,k+1},\dots,\hat{y}^{NT}_{ref,k+h}\}$ and $\{\hat{y}^{NT}_{k+1},\dots,\hat{y}^{NT}_{k+h}\}$, respectively, and the next $hN$ disturbances $\{\hat{d}^T_{k+N},\dots,\hat{d}^T_{k+hN-1}\}$, the set of future control actions $\{\hat{u}^T_{k+N},\dots,\hat{u}^T_{k+hN-1}\}$ can be obtained.

Notice that the output of $G_1^{NT}(z^N)$ (i.e, $u_{1,k}^{NT}$) is expanded $[u_{1,k}^{NT}]^T$ before being injected to the digital hold $H^{NT,T}(z)$. The expand operation implies to fill the slow-rate signal with zeros at the fast-rate instants (more details can be found in \cite{salt2005mbm}). 

The digital hold $H^{NT,T}(z)$ retains the last injected value of the slow-rate controller output signal $u_{1,k}^{NT}$, which implies that the sub-controller output $u_{1,k}^{NT}$ is repeated $N$ times. Hence, the transfer function of the digital hold can be expressed as:
$$
H^{NT,T}(z)=\frac{1-z^{-N}}{1-z^{-1}}
$$

The dual-rate control design method can be performed from the desired transfer function of the continuous-time closed-loop system $M(s)$. Let us consider $M^T(z)$ and $M^{NT}(z^N)$ as the ZOH-based discretization of $M(s)$ at two different sampling periods: $T$ and $NT$, respectively. Then, the subcontrollers $G_1^{NT}(z^N)$ and $G_2^{T}(z)$ will be designed as follows \cite{salt2005mbm}:
\begin{eqnarray}
G_1^{NT}(z^N) & = & \frac{1}{1-M^{NT}(z^N)}  \label{lento}
\end{eqnarray}
and
\begin{equation}
G_2^T(z)=\frac{M^T(z) }{G_p^T(z)} \label{rapido}
\end{equation}

where $G_p^T(z)=C_p \left(z I - A_p\right)^{-1} B_p$ is the discrete-time transfer function of the nominal system \eqref{eq:model} at period $T$ under ZOH conditions.

\section{Stability analysis}\label{sec:analysis}
Before proceeding with the stability analysis, an equivalent interconnected state-space representation of the closed-loop system \eqref{eq:model} and the proposed NCS in Section \ref{sec:control} is obtained through Section \ref{sec:ssdes} and \ref{sec:ssmod}. The closed-loop model \eqref{eq:intercon} allows us to further address the stability analysis in the framework of LMI via Lyapunov method in Section \ref{sec:stana}.

\subsection{State-space description}\label{sec:ssdes}
Since each device operates at a different period, the plant model and the different components of the control system presented in Section \ref{sec:control} are first converted at fast rate using lifting techniques. In this way, an equivalent interconnected representation for the closed-loop system is found in \eqref{eq:intercon} by  including the above defined event-triggered protocols and time-varying model uncertainties. For this purpose, input/output approaches have been suitably applied to embed them into a single norm-bounded uncertain system (namely $\Delta$ in \eqref{eq:intercon}). Next subsections give a detailed description of how each subsystem model has been obtained to further integrate them into \eqref{eq:intercon}.

\subsubsection{Plant model}

To deal with the event-triggered protocol for transmission of control actions defined in Section \ref{sec:evtact} and model uncertainties \eqref{eq:uncertt}, let us define the following new inputs:
\begin{align}\label{eq:rhou}
&\rho_{u,k}^{T} =  u_k^{T} - \tilde{u}_k^{T}, \\ \nonumber
&w_{\Delta,k}^T = \Delta_k y_{\Delta,k}^T
\end{align}
where $y_{\Delta,k}^T = H_A x_k^T + H_{p,B} u_k^T$.

From the above definition and \eqref{eq:uncertt}, we have that $u_k^{T} = \tilde{u}_k^{T} + \rho_{u,k}^{T}$ and $\Delta_{A,k} + \Delta_{B,k} = E w_{\Delta,k}^T$. Then, \eqref{eq:plantd} can be rewritten as the interconnected system:
\begin{align}\label{eq:plantd_augm}
&x_{k+1}^T = A x_{k}^T + B \tilde{u}_k^T + B \rho_{u,k}^T + B_{\omega} \omega_k^T + \delta_{\Delta} E w_{\Delta,k}^T, \\ \nonumber
&y_{\Delta,k} ^T= H_A x_k^T + H_{p,B} \tilde{u}_k^T + H_{p,B} \rho_{u,k}^T, \\ \nonumber
&y_k^T = C x_k^T + v_k^T
\end{align}
where $w_{\Delta,k}^T = \Delta_k y_{\Delta,k}^T$.

\subsubsection{Time-Varying Dual-Rate Kalman Filter}

To cope with the event-triggered protocol for transmission of measurement data defined in Section \ref{sec:evtmes}, let us introduce the following input:
\begin{align}\label{eq:rhoy}
\rho_{y,k}^{NT} =\tilde{y}_k^{NT} - y_k^{NT}
\end{align}
From the above definition, we have that $\tilde{y}_k^{NT} = y_k^{NT} +\rho_{y,k}^{NT}$. The TVDRKF \eqref{eq:kalm} and \eqref{eq:predy} can be expressed in augmented form at fast period $T$ as:
\begin{align}\label{eq:tvdrkf}
&\bar{\hat{x}}_{k+1}^T = \mathcal{A}_{\bar{N}} \bar{\hat{x}}_{k}^T + \mathcal{B}_{\bar{N}} \bar{\tilde{u}}_{k}^T + \left( \mathcal{K}_{\bar{N}} C \right) x_{k}^T +  \left(  \mathcal{K}_{\bar{N}} C \right) \rho_{y,k}^T + \mathcal{K}_{\bar{N}} v_k^T
\end{align}
where
\begin{align}\label{eq:statekal}
&\bar{\hat{x}}_{k+1}^T = \begin{bmatrix} \left(\hat{x}_{k|k}^{T}\right)^{'} & \ \left(\hat{x}_{k-1|k-\bar{N}}^{T}\right)^{'}  & \cdots &   \left(\hat{x}_{k-\bar{N}+1|k-\bar{N}}^{T}\right)^{'}  \end{bmatrix}', \\ \nonumber
&\bar{\hat{x}}_{k}^T = \begin{bmatrix} \left(\hat{x}_{k-1|k-\bar{N}}^{T}\right)^{'}  & \ \cdots &   \left(\hat{x}_{k-\bar{N}|k-\bar{N}}^{T}\right)^{'}  \end{bmatrix}', \\ \nonumber
&\bar{\tilde{u}}_{k}^T = \begin{bmatrix} \left(\tilde{u}_{k-1}^{T}\right)^{'}  & \ \left(\tilde{u}_{k-2}^{T}\right)^{'}   & \cdots & \left(\tilde{u}_{k-\bar{N}}^{T}\right)^{'}  \end{bmatrix}',
\end{align}
and
\begin{align}
&\mathcal{A}_{\bar{N}} = \begin{bmatrix} 0_n & \ 0 & \  A^{\bar{N}}-K(\bar{N}) C  A^{\bar{N}}  \\
I_n & \ 0 & \ 0 \\
0 & \ I_{(\bar{N}-2) n} & 0_{(\bar{N}-2) n \times n}
\end{bmatrix}, \\ \nonumber
& \mathcal{B}_{\bar{N}} = \begin{bmatrix} B_{\bar{N}} \\
0_{(\bar{N}-1) n \times \bar{N} m}
\end{bmatrix}, \\ \nonumber
&B_{\bar{N}} = \begin{bmatrix}  B & \ A B & \cdots & \ A^{\bar{N}-1} B \end{bmatrix}, \\ \nonumber
&\mathcal{K}_{\bar{N}} = \begin{bmatrix}K^{'}(\bar{N}) & \ 0_{q \times (\bar{N}-1) n}  \end{bmatrix}^{'},
\end{align}
where $n = n_p + r$
\subsubsection{Slow-rate controller}

The state-space model of the slow-rate controller $G_1^{NT}(z^N)$ can be expressed as:
\begin{align}\label{eq:slow}
&\xi_{k+1}^{NT}  = A_{\xi} \xi_{k}^{NT} + B_{\xi} e_{k}^{NT}, \\ \nonumber
&u_{1,k}^{NT} = C_{\xi} \xi_{k}^{NT} + D_{\xi} e_{k}^{NT}
\end{align}
where $e_k^{NT} = y_{ref,k}^{NT} - \hat{y}_k^{NT}$, $\hat{y}_k^{NT} = C \hat{x}_{k|k}^{NT}$, $\xi_k \in \mathcal{R}^{n_{\xi}}$ is the state vector of the slow-rate controller, and $A_{\xi} \in \mathcal{R}^{n_{\xi} \times n_{\xi}}$, $B_{\xi} \in \mathcal{R}^{n_{\xi} \times q}$, $C_{\xi} \in \mathcal{R}^{q_{\xi} \times n_{\xi}}$, $D_{\xi} \in \mathcal{R}^{q_{\xi} \times q}$   are state-space matrices satisfying $G_1^{NT}(z) = C_{\xi} \left(zI-A_{\xi}\right)^{-1} B_{\xi} + D_{\xi}$.

From the expand operator it can be deduced that $C_{\xi} \xi_{k}^{NT}=\bar{C}_{\xi} \bar{\xi}_{k}^{T}$, where  $
\bar{C}_{\xi} =  \begin{bmatrix} 0 & \ C_{\xi} \end{bmatrix}$
and
\begin{align}\label{eq:statefast}
&\bar{\xi}_{k}^{T} = \begin{bmatrix} \left(\xi_{k-1}^{T}\right)^{'} & \ \left(\xi_{k-2}^{T}\right)^{'} & \ \cdots &   \left(\xi_{k-N}^{T}\right)^{'} \end{bmatrix}'.
\end{align}
Applying Remark \ref{remark:aummm}, the slow-rate controller together with the expand operator can equivalently be represented at fast period $T$ as:
\begin{align}\label{eq:slow_h2}
&\bar{\xi}_{k+N}^{T} = \bar{A}_{\xi} \bar{\xi}_{k}^{T} + \bar{B}_{\xi} e_k^{T}, \\ \nonumber
&u_{1,k}^{T} = \bar{C}_{\xi} \bar{\xi}_{k}^{T} + D_{\xi} e_k^{T}
\end{align}
where
\begin{align}
\bar{A}_{\xi} = \begin{bmatrix} 0 & \ A_{\xi} \\ I_{(N-1)n_{\xi}} & \ 0 \end{bmatrix}, \quad \bar{B}_{\xi} =  \begin{bmatrix} B_{\xi} \\ 0 \end{bmatrix}
\end{align}

\subsubsection{Fast-rate controller:}
The state-space model of the fast-rate controller $G_2^T(z)$ with anti-disturbance can be expressed as:
\begin{align}\label{eq:fast}
&\eta_{k+1}^T = A_{\eta} \eta_{k}^T + B_{\eta} u_{1,k}^{T}, \\ \nonumber
&\tilde{u}_k^T = C_{\eta} \eta_{k}^T + D_{\eta} u_{1,k}^{T} - \hat{d}_k^T
\end{align}
where $u_{1,k}^{T}$ is the input of the fast-rate controller, $\eta_k \in \mathcal{R}^{n_{\eta}}$ is the state vector of the fast-rate controller and $A_{\eta} \in \mathcal{R}^{n_{\eta} \times n_{\eta}}$, $B_{\eta} \in \mathcal{R}^{n_{\eta} \times q_{\xi}}$, $C_{\eta} \in \mathcal{R}^{m \times n_{\eta}}$, $D_{\eta} \in \mathcal{R}^{m \times q_{\xi}}$ are state-space matrices satisfying $G_2^T(z) = C_{\eta} \left(zI-A_{\eta}\right)^{-1} B_{\eta} + D_{\eta}$.

\subsection{Closed-loop system}\label{sec:ssmod}
Recalling the definition of tracking error $e_k^{NT} = y_{ref,k}^{NT} - \hat{y}_k^{NT}$ and the definition of $ \bar{\hat{x}}_k^{NT} $ in  \eqref{eq:statekal}, we have that
\begin{align}\label{eq:entrada}
\bar{e}_k^{NT}  = \bar{y}_{ref,k}^{NT} - \bar{C}_y \bar{\hat{x}}_k^{NT}
\end{align}
where $\bar{y}_{ref,k}^{NT} = \begin{bmatrix} \left(y_{ref,k-1}^{NT}\right)^{'} & \cdots, \left(y_{ref,k-h}^{NT}\right)^{'} \end{bmatrix}^{'}$ and $\bar{C}_y = \begin{bmatrix} C & \quad 0_{q \times (\bar{N}-1) n} \end{bmatrix}$.

Hence, from \eqref{eq:plantd_augm}, \eqref{eq:tvdrkf}, \eqref{eq:slow_h2}, \eqref{eq:fast} and \eqref{eq:entrada}, together with the model uncertainties given in \eqref{eq:uncertt}, the closed-loop control system can be represented as the interconnected system at period $T$:
\begin{align}\label{eq:intercon}
&\mathcal{S} : \begin{cases} \bar{\phi}_{k+1}^T = A_{\phi} \bar{\phi}_k^T + B_{\phi} \bar{\rho}_k^T + B_{\omega,\phi} \bar{\omega}_k^T + \delta_{\Delta} E_{\phi} w_{\Delta,k}^T +B_{ref,\phi} \bar{y}_{ref,k}^T & \\
y_{\Delta,k}^T = H_{\phi} \bar{\phi}_k^T + H_{p,B} \bar{\rho}_k^T  & \\
y_k^T = C_{y,\phi} \bar{\phi}_k^T  + C_{\omega,\phi} \bar{\omega}_k^T & \end{cases}, \\ \nonumber
&\Delta : \begin{cases} w_{\Delta,k}^T = \Delta_k y_{\Delta,k}^T  \end{cases}
\end{align}
where
\begin{align}\label{eq:estadosaumentaos}
&\bar{\phi}_k^T = \begin{bmatrix} \left(x_k^{T}\right)^{'} & \ \left(\bar{\tilde{u}}_k^{T}\right)^{'} & \ \left(\bar{\hat{x}}_k^{T}\right)^{'} & \ \left(\bar{\xi}_k^{T}\right)^{'} & \ \left(\eta_k^{T}\right)^{'} \end{bmatrix}^{'}, \\ \nonumber
&\bar{\rho}_k^T = \begin{bmatrix} \left(\rho_{u,k}^{T}\right)^{'} & \ \left(\rho_{y,k}^{T}\right)^{'}\end{bmatrix}^{'}, \\ \nonumber
&\bar{\omega}_k^T = \begin{bmatrix} \left(\omega_k^{T}\right)^{'} & \ \left(v_{k}^{T}\right)^{'}\end{bmatrix}^{'}
\end{align}
\begin{align}\label{eq:intercon_matrices}
&A_{\phi}  = \begin{bmatrix}
A & \ 0 & \ -B D_{\eta} D_{\xi} \bar{C}_y - B \bar{\mathcal{C}}_d & \ B D_{\eta} \bar{C}_{\xi} & \ B C_{\eta} \\
0 & \ \Gamma_1 & \ \Gamma_2 & \ \Gamma_3 & \ \Gamma_4 \\
\mathcal{K}_{\bar{N}} C & \ (\mathcal{B}_{\bar{N}}-\mathcal{K}_{\bar{N}} C B_{\bar{N}})  & \ \mathcal{A}_{\bar{N}} & \ 0 & \ 0 \\
0 & \ 0 & \ -\bar{B}_{\xi} \bar{C}_y & \ \bar{A}_{\xi} & \ 0 \\
0 & \ 0 & \  -B_{\eta} D_{\xi} \bar{C}_y  & \ B_{\eta} \bar{C}_{\xi} & \ A_{\eta}
\end{bmatrix}, \\ \nonumber
&B_{\phi}= \begin{bmatrix} B& \ 0 \\ 0 & \ 0 \\ 0 & \  \mathcal{K}_{\bar{N}} \mathcal{C}_{\bar{N}}  \\ 0  & \ 0 \\ 0  & \ 0 \end{bmatrix},\quad B_{ref,\phi}= \begin{bmatrix} B D_{\eta} D_{\xi} \\ \Gamma_5 \\ 0 \\ \bar{B}_{\xi} \\ B_{\eta} D_{\xi} \end{bmatrix}, \ B_{\omega,\phi} = \begin{bmatrix} B_{\omega} & \ 0 \\ 0 & \ 0 \\ 0 & \ 0 \\ 0 & \ 0 \\ 0 & \ 0 \end{bmatrix} \ E_{\phi} = \begin{bmatrix}E \\ 0 \\ 0 \\ 0 \\ 0 \end{bmatrix}, \\ \nonumber
&C_{y,\phi} = \begin{bmatrix} C & \ 0   & \ 0  & \ 0  & \ 0  \end{bmatrix}, \\ \nonumber
&\bar{\mathcal{C}}_{d} = \begin{bmatrix} \mathcal{C}_d & \ 0   & \ 0  & \ 0  & \ 0  \end{bmatrix}, \\ \nonumber
&C_{\omega,\phi}  = \begin{bmatrix} 0_q & \ I_q \end{bmatrix}, \\ \nonumber
&H_{\phi} = \begin{bmatrix}H_A & \ 0_{l_2 \times \bar{N}m} & \ -H_{p,B} D_{\eta} D_{\xi} \bar{C}_y & \ H_{p,B} D_{\eta} \bar{C}_{\xi} & \ H_{p,B} C_{\eta}
\end{bmatrix}, 
\end{align}
and
\begin{align}
&\Gamma_1 = \begin{bmatrix} 0 & \ 0 \\ I_{(\bar{N} - 1)m} & 0 \end{bmatrix}, \qquad
\Gamma_2 = \begin{bmatrix} - D_{\eta} D_{\xi} \bar{C}_y - \bar{\mathcal{C}}_d \\ 0_{(\bar{N} - 1)m \times \bar{N} n} \end{bmatrix}, \\ \nonumber
&\Gamma_3 = \begin{bmatrix} D_{\eta} \bar{C}_{\xi} \\ 0_{(\bar{N} - 1)m \times N n_{\xi} }  \end{bmatrix}, \qquad
\Gamma_4 = \begin{bmatrix} C_{\eta} \\ 0_{(\bar{N} - 1)m \times n_{\eta}}  \end{bmatrix}, \qquad
\Gamma_5 = \begin{bmatrix} D_{\eta} D_{\xi} \\ 0_{(\bar{N} - 1)m \times q}  \end{bmatrix}
\end{align}

\subsection{Robust stability analysis of the proposed NCS design}\label{sec:stana}
This section gives a sufficient condition to ascertain the robust stability of the closed-loop control system \eqref{eq:intercon} against time-varying model uncertainties \eqref{eq:uncertt}.
\begin{theorem}\label{teorema1}
	Given scalars $\sigma_u, \sigma_y \geq 0$, system \eqref{eq:intercon} is robustly stable if there exist symmetric matrices $P \in \mathcal{R}^{\bar{n}}, \Omega_u \in \mathcal{R}^m, \Omega_y \in \mathcal{R}^q > 0$ with $\bar{n}=n+h N (m + n) +N n_{\xi}+n_{\eta}$ and a scalar $\varepsilon>0$ such that the following LMI holds:
	\begin{align}\label{eq:lmitoral}
	\begin{bmatrix}
	-P & 0 & \ 0 & \ \ A_{\phi}' P & \ C_{\phi}' \bar{\Omega} &  \  H_{\phi}' \\
	(*) & \ -\bar{\Omega} & \ 0 & \ B_{\phi}' P & \ 0 & \ 0 \\
	(*) & \ (*) & \ -\varepsilon I_{l_1} & \ E_{\phi}' P  & \ 0  & \ 0 \\
	(*) & \ (*) & \ (*) & \ -P & \ 0 &  \ 0 \\
	(*) & \ (*) &  \ (*) & \ (*) &  \ -\bar{\Omega}  & \ 0 \\
	(*) & \ (*) & \ (*) & \ (*) &  \ (*) & \ -I_{l_2}
	\end{bmatrix} < 0
	\end{align}
	where $A_{\phi}, B_{\phi}, E_{\phi} ,H_{\phi}, $ are defined in \eqref{eq:intercon_matrices}, and
	\begin{align}\label{eq:cfii}
	&\bar{\Omega} = diag\left(\Omega_u, \ \Omega_y \right), \\ \nonumber
	&C_{\phi} = \begin{bmatrix} 0 & \ 0 \ & - \sigma_u D_{\eta} D_{\xi} \bar{C}_y & \sigma_u D_{\eta} \bar{C}_{\xi} & \sigma_u C_{\eta} \\
	\sigma_y C & \ 0   & \ 0  & \ 0  & \ 0  \end{bmatrix}
	\end{align}
	Moreover, a level of robustness $\delta_{\Delta}=\varepsilon^{-1/2}$ is guaranteed, where $\delta_{\Delta}$ is defined in \eqref{eq:uncertt}.
\end{theorem}
\begin{proof}
	For stability analysis, consider null values for reference signal and external disturbance (i.e, $ \bar{y}_{ref,k}^T  = 0$ and $\bar{\omega}_k^T = 0$). Then, the forward system $\mathcal{S}$ in \eqref{eq:intercon} can be written as:
	\begin{align}\label{eq:intercon_clean}
	\bar{\phi}_{k+1}^T = A_{\phi} \bar{\phi}_k^T + B_{\phi} \bar{\rho}_k^T +  \delta_{\Delta} E_{\phi} w_{\Delta,k}^{T} 
	\end{align}
	Consider the Lyapunov function $V_k = \left(\phi_k^{T}\right)^{'} P \phi_k^T$. Taking into account \eqref{eq:intercon_clean}, the forward difference operator $\Delta V_k = V_{k+1} - V_k$ renders:
	\begin{align}\label{eq:DV}
	&\Delta V_k = \left(\phi_{k+1}^{T}\right)^{'} P \phi_{k+1}^T - \left( \phi_{k}^{T} \right)^{'} P \phi_k^T \\ \nonumber
&= \left( \phi_{k}^{T} \right)^{'} \left( A_{\phi}' P A_{\phi} - P  \right) \phi_k^{T} + 2 \left( \phi_{k}^{T} \right)^{'} A_{\phi}' P B_{\phi} \bar{\rho}_k^T  \\ \nonumber
	&+ 2 \delta_{\Delta} \left( \phi_{k}^{T} \right)^{'} A_{\phi}' P E_{\phi} w_{\Delta,k}^T +   \left( \bar{\rho}_{k}^{T} \right)^{'} B_{\phi}' P B_{\phi} \bar{\rho}_k^{T} \\ \nonumber
& + 2 \delta_{\Delta} \left( \bar{\rho}_{k}^{T} \right)^{'} B_{\phi}' P E_{\phi} w_{\Delta,k}^T + \delta_{\Delta}^2 \left( w_{\Delta,k}^{T} \right)^{'} E_{\phi}' P E_{\phi} w_{\Delta,k}^{T}
	\end{align}
	Pre-and post-multiplying \eqref{eq:lmitoral} by diag$\left(I, I, \delta_{\Delta} I, P^{-1}, \bar{\Omega}^{-1}, I\right)$ and further applying Schur Complement in the third and fourth rows and columns, it is easy to deduce that 
	\begin{align} \nonumber
	&\begin{bmatrix} \phi_{k}^{T}\\ \bar{\rho}_{k}^{T} \\ w_{\Delta,k}^{T} \end{bmatrix}' \begin{bmatrix} \begin{matrix} A_{\phi}' P A_{\phi} - P \\ + C_{\phi}' \bar{\Omega} C_{\phi} + H_{\phi}' H_{\phi} \end{matrix} & A_{\phi}' P B_{\phi}  & \delta_{\Delta} A_{\phi}' P E_{\phi}  \\
	(*) & B_{\phi}' P B_{\phi} - \bar{\Omega} & \delta_{\Delta} B_{\phi}' P E_{\phi} \\
	(*) & (*) & \delta_{\Delta}^2 E_{\phi}' P E_{\phi} -I_{l_1}\end{bmatrix} \begin{bmatrix} \phi_k^{T} \\ \bar{\rho}_k^{T} \\ w_{\Delta,k}^{T}\end{bmatrix} < 0
	\end{align}
	is true $\forall \phi_k^T, \bar{\rho}_k^T,  w_{\Delta,k}^T \neq 0$ if and only if \eqref{eq:lmitoral} holds. 
	

	From \eqref{eq:y} and \eqref{eq:cond2} it can be seen that:
	\begin{align}\label{conddddy}
	(y_k^{T} - \tilde{y}_k^{T})^{'} \Omega_y (y_k^{T} - \tilde{y}_k^{T}) \leq \sigma_y^2 \left( y_k^{T} \right)^{'} \Omega_y y_k^{T} + \delta_y
	\end{align}
	Analogously, from \eqref{eq:cond1}  and \eqref{eq:u}, the following inequality is true
	\begin{align}\label{conddddu}
	(u_k^T - \tilde{u}_k^T)^{'} \Omega_u (u_k^T - \tilde{u}_k^T) \leq \sigma_u^2 \left( u_k^{T} \right)^{'} \Omega_u u_k^T + \delta_u
	\end{align}
	Applying the definition of $\rho_{u,k}^T$, $\rho_{y,k}^T$ in \eqref{eq:rhou}, \eqref{eq:rhoy} respectively, the above inequalities \eqref{conddddu},  \eqref{conddddy} can be rewritten as:
	\begin{align}\label{eq:clue}
	&\left( \rho_{u,k}^{T} \right)^{'} \Omega_u \rho_{u,k}^T \leq  \sigma_u^2 \left( u_k^{T} \right)^{'} \Omega_u u_k^T + \delta_u, \\ \nonumber
	&\left( \rho_{y,k}^{T} \right)^{'} \Omega_y \rho_{y,k}^T \leq  \sigma_y^2 \left( y_k^{T} \right)^{'}  \Omega_y y_k^T +  \delta_y
	\end{align}
	The inequalities \eqref{eq:clue} can be formulated in compact form as:
	\begin{align}\label{eq:ccevt}
	&\left( \bar{\rho}_{k}^{T} \right)^{'} \bar{\Omega} \bar{\rho}_{k}^T \leq  \left( \bar{y}_{\phi,k}^{T} \right)^{'}  \bar{\Omega} \bar{y}_{\phi,k}^T + \delta
	\end{align}
	where $\bar{\Omega} = diag\left(\Omega_u, \ \Omega_y \right)$, $\delta=\delta_u +\delta_y$ and 
	\begin{align}
	\bar{y}_{\phi,k}^T = C_{\phi} \bar{\phi}_k^T
	\end{align}
	where $C_{\phi}$ is defined in \eqref{eq:cfii} and $\bar{\phi}_k^T$ is defined in \eqref{eq:estadosaumentaos}. From the definition of $\Delta_k$ in \eqref{eq:uncertt}, it can be deduced that
	\begin{align}\label{eq:ccdel}
	\left( y_{\Delta,k}^{T}\right)^{'} y_{\Delta,k}^T \leq \left( w_{\Delta,k}^{T} \right)^{'} w_{\Delta,k}^{T}
	\end{align}
	By considering $\delta=0$, the fulfilment of the condition $\bar{J} = \sum_{k=0}^{\infty} \left(\Delta V_k  + J_k \right)< 0$ with
	\begin{align}
	&J_k = \left( \bar{\rho}_{k}^{T} \right)^{'} \bar{\Omega} \bar{\rho}_{k}^T -  \left( \bar{y}_{\phi,k}^{T} \right)^{'}  \bar{\Omega} \bar{y}_{\phi,k}^{T}  + \left(y_{\Delta,k}^{T}\right)^{'} y_{\Delta,k}^T -  \left( w_{\Delta,k}^{T} \right)^{'} w_{\Delta,k}^{T}
	\end{align}
	implies: (i) the negativeness of the forward difference: $\Delta V_k < 0$ and, (ii) the fulfillment of the inequalities \eqref{eq:ccevt} and \eqref{eq:ccdel} under zero initial conditions for the augmented state variable $\bar{\phi}_k^T$ in \eqref{eq:estadosaumentaos}. The former condition proves the stability of the closed-loop system, whereas the latter condition proves: (i) the robustness stability under event-triggered condition for some $\sigma_u, \sigma_y>0$, and (ii) robustness against model uncertainties given by $\left(y_{\Delta,k}^{T}\right)^{'} y_{\Delta,k}^T \leq \left( w_{\Delta,k}^{T} \right)^{'} w_{\Delta,k}^{T}$. Finally, by considering $\delta >0$, the above condition renders $\bar{J} < \delta$ when the system state is close to the equilibrium point, and $\bar{J} < 0$ otherwise. Hence, the steady-state error remains bounded inside a ball whose radius is proportional to $\delta$.
\end{proof}

\begin{remark}\label{rem_opt}
	The maximum level of robustness of the closed-loop system \eqref{eq:intercon} can be determined by solving the following convex optimization problem:
	\begin{align}
	\text{min } \varepsilon \text{ s.t. LMI \eqref{eq:lmitoral}}
	\end{align}
	where the upper bound for model uncertainties defined in \eqref{eq:uncertt} can be obtained as $\delta_{\Delta} = \varepsilon^{-1/2}$.
\end{remark}

\begin{remark}
The robust stability algorithm given in Theorem \ref{teorema1} is based on the quadratic Lyapunov function $V_k = \left(\phi_k^{T}\right)^{'} P \phi_k^T$. The advantage of this choice is that conservatism is reduced to the greatest extent since no loss of generality is introduced in the Lyapunov function. Nevertheless, the drawback is that computational complexity is proportional to the number of consecutive packet dropouts $h$, as can be deduced from the number of decision variables (NoV) and size of LMI \eqref{eq:lmitoral}, which are respectively $\frac{1}{2} \bar{n} \left(\bar{n} + 1\right) + \frac{1}{2} m \left(m + 1\right) + \frac{1}{2} q \left(q + 1\right) + 1$ and $2 \bar{n} + m + q + l_1 + l_2$, where $\bar{n}=n+h N (m + n) +N n_{\xi}+n_{\eta}$. 
\end{remark}

\section{Application to a UGV}\label{sec:UGV}
In this section, the control solution proposed in this work is applied to a two-wheel UGV tracking control. In order to carry out this kind of control, the reference generator depicted in Fig. \ref{fig:ControlScheme} needs to include a path tracking algorithm. In this case, Pure Pursuit is chosen (introduced in Section \ref{sec:PurePursuit}). The UGV and the main application data are presented in Section \ref{sec:Data}, where the closed-loop stability analysis is also performed. Simulation results, together with their experimental validation, are shown in Section \ref{sec:SimExp}, where some cost indexes (presented in section \ref{sec:Costfunc}) are used to better quantify the benefits of the control approach, compared to other options.

\subsection{Pure Pursuit path tracking algorithm}\label{sec:PurePursuit}

The Pure Pursuit path tracking algorithm is in charge of computing the velocity and turning conditions so that the UGV can follow a prescribed trajectory \cite{lundgren2003path}. More concretely, by means of this algorithm, and using an odometry technique, the UGV is able to infer its current position from the prescribed kinematic reference $(X_{ref},Y_{ref},\theta_{ref})^{NT}_k$ and the plant output estimate generated by the TVDRKF at period $NT$,
$(\hat{w}_r,\hat{w}_l)^{NT}_k\equiv\hat{y}^{NT}_k$, being $(\hat{w}_r,\hat{w}_l)^{NT}_k$ the rotational velocity for the right and left wheel motors, respectively. Then, with the aim of properly reaching
the next point of the desired trajectory, the algorithm generates the dynamic reference based on the
rotational velocity for both wheels, i.e.,
$(w_{r,ref},w_{l,ref})^{NT}_k\equiv
y^{NT}_{ref,k}$. 

The UGV path tracking is composed of a set of $h$ future dynamic references,
$\{y^{NT}_{ref,k},y^{NT}_{ref,k+1},\dots,y^{NT}_{ref,k+h}\}$. In order to establish these references, the reference generator provides in advance the Pure Pursuit
algorithm with the sequence of $h$-step ahead kinematic references:
$$\{(X_{ref},Y_{ref},\theta_{ref})^{NT}_{k},
(X_{ref},Y_{ref},\theta_{ref})^{NT}_{k+1},\dots,(X_{ref},Y_{ref},\theta_{ref})^{NT}_{k+h}\}
$$
The main steps of the Pure Pursuit algorithm are (more details can be found in \cite{alcaina2019energy}):
\begin{enumerate} 
	\item 
	 Calculation of rotational and linear velocities ($\omega_k^{NT}$ and $v_k^{NT}$, respectively) from $\hat{w}^{NT}_{r,k}$ and $\hat{w}^{NT}_{l,k}$:
	\begin{eqnarray}
	v_{r,k}^{NT} & = & \hat{w}^{NT}_{r,k}r \\
	v_{l,k}^{NT} & = & \hat{w}^{NT}_{l,k}r \\
	v_k^{NT} & =& \frac{v_{r,k}^{NT}+v_{l,k}^{NT}}{2} \\
	\omega_k^{NT} & =& \frac{v_{r,k}^{NT}-v_{l,k}^{NT}}{l}
	\end{eqnarray}
	being $r$ the wheel radius, 
	and $l$ the distance between the two wheels.
	\item UGV position and orientation computation in the time period $NT$:
	\begin{eqnarray}\label{kinematic}
	X^{NT}_k & = & X^{NT}_{k-1}+v^{NT}_{k}NT\cos(\theta^{NT}_{k-1}+\omega_k^{NT}NT) \label{kinematic1} \\
	Y^{NT}_k & = & Y^{NT}_{k-1}+v_k^{NT}NT\sin(\theta^{NT}_{k-1}+\omega_k^{NT}NT) \label{kinematic2}\\
	\theta^{NT}_k & = & \theta^{NT}_{k-1}+\omega_k^{NT}NT
	\end{eqnarray}
	$\textrm{for} \ k \in \mathbb{N}_{\geq1}$, where $(X,Y,\theta)^{NT}_0$ is the initial position and orientation.
	\item Generation of the future reference for the UGV, $(X_{ref},Y_{ref})^{NT}_k$, from the Look Ahead Distance (LAD) and the prescribed kinematic reference.
	\item Computation of control law $\bar{k}$ at period $NT$: 
	\begin{equation}
	\bar{k}=2\frac{(Y_{ref,k}^{NT}-Y_k^{NT})\cos(\theta_k^{NT})-(X_{ref,k}^{NT}-X_k^{NT})\sin (\theta_k^{NT})
	}{(X_{ref,k}^{NT}-X_k^{NT})^2+(Y_{ref,k}^{NT}-Y_k^{NT})^2} \label{kPP}
	\end{equation}
	As a result, and given a desired linear velocity $v^{NT}_{ref,k}$, 
	the rotational velocity reference $\omega^{NT}_{ref,k}$ and the dynamic references $(w_{r,ref},w_{l,ref})^{NT}_k$ are calculated as follows:
	\begin{equation}
	\omega^{NT}_{ref,k}=v^{NT}_{ref,k}\bar{k} \label{omega}
	\end{equation}
	\begin{eqnarray}
	(w_{r,ref})_k^{NT}=\frac{v^{NT}_{ref,k}+\omega^{NT}_{ref,k}b}{r} \\
	(w_{l,ref})_k^{NT}=\frac{v^{NT}_{ref,k}-\omega^{NT}_{ref,k}b}{r}
	\end{eqnarray}
	where $b$ is half of the distance between the two wheels.
\end{enumerate}

\begin{figure}[H]
	\centering
	\includegraphics[width=13.0cm, height=8.0cm]{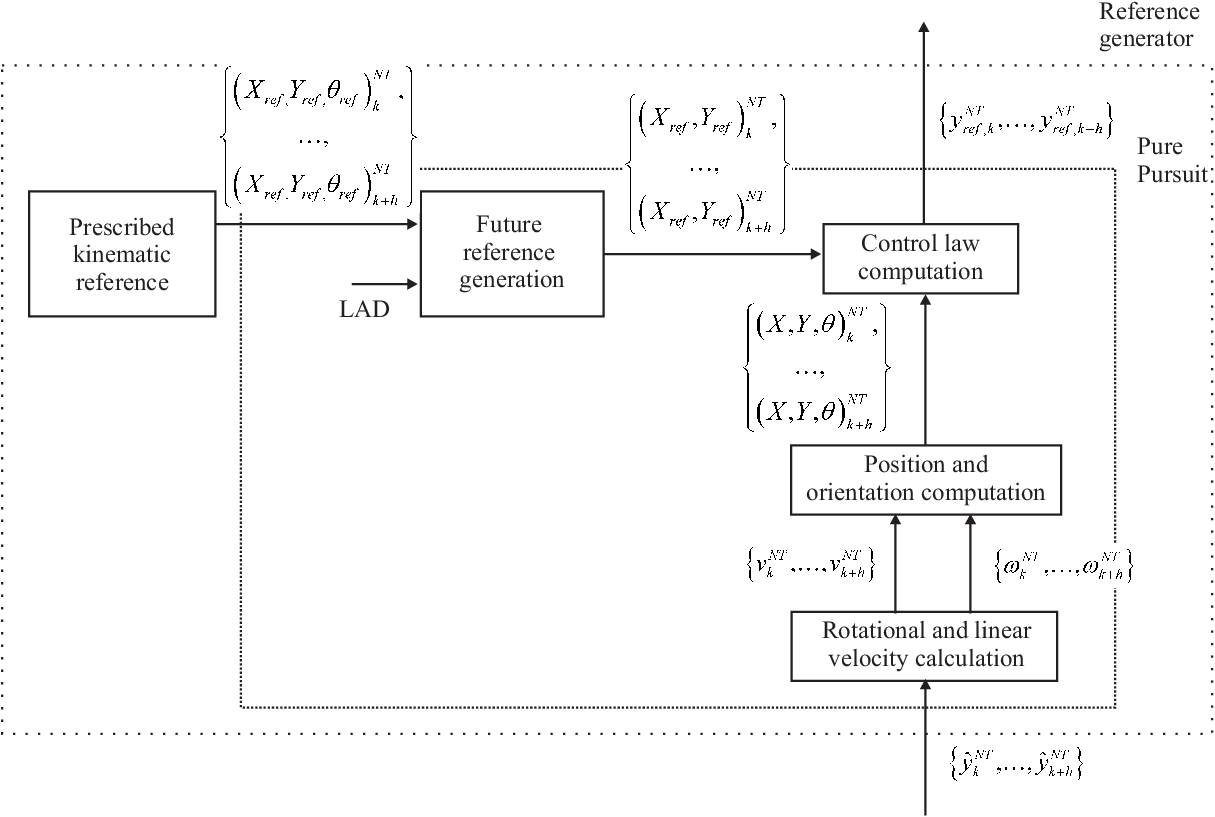}
	\caption{Structure of the reference generator, including Pure Pursuit}
	\label{fig:Pure}
\end{figure}

The algorithm is repeated for the $h$-step ahead predictions. Figure \ref{fig:Pure} illustrates the different algorithm stages.

\subsection{Cost indexes for control performance and resource usage}\label{sec:Costfunc}
In this section, some cost indexes are presented to better quantify control performance and resource
usage. Some comparisons will be carried out in Section \ref{sec:SimExp}, focusing on revealing the benefits of the proposed energy-efficient event-triggered control solution versus the conventional time-triggered one. 

Three cost indexes will be used to analyze control performance (see \cite{alcaina2019energy} for details):

\begin{itemize}
	\item $J_1$, which is based on the $\ell_2$-norm:
	\begin{equation}\label{J1}
	J_1=\sum_{k=1}^{l} \min_{1\leq k' \leq l}  \sqrt{(X_k^{NT}-X_{ref,k'}^{NT})^2+(Y_k^{NT}-Y_{ref,k'}^{NT})^2}
	\end{equation}
	where $l$ is the number of iterations to reach the final destination, 
	and $(X_{ref},Y_{ref})_{k'}^{NT}$ is the nearest kinematic position reference to the current UGV position.
	
	\item $J_2$, which is based on the $\ell_\infty$-norm:
	\begin{equation}\label{J2}
	J_2=\max_{1\leq k \leq l} \left\{\min_{1\leq k' \leq l}  \sqrt{(X_k^{NT}-X_{ref,k'}^{NT})^2+(Y_k^{NT}-Y_{ref,k'}^{NT})^2}\right\}
	\end{equation}
	
	\item $J_3$, which measures the total amount of time (in seconds) required to reach the final destination: 
	\begin{equation}\label{J3}
	J_3=lNT
	\end{equation}
\end{itemize}

As in \cite{alcaina2019energy}, to analyze resource saving, the cost index $J_4$ measures (in \%) the reduction of the number of packets transmitted by the proposed solution, $NoT_\textrm{EEC}$, versus the conventional one, $NoT_\textrm{TTC}$: 
\begin{equation}\label{J4}
J_4=\frac{NoT_\textrm{EEC}}{NoT_\textrm{TTC}}\cdot100 \%.
\end{equation}

\subsection{Robust stability analysis}\label{sec:Data}
Consider a plant system consisting in the two-wheel UGV. The system model for both wheel motors is described by means of the following transfer function \cite{alcaina2019energy}:
\begin{align}\label{eq:Gpppp}
G_p(s) = \frac{0.1276}{0.1235s + 1}
\end{align}
where the output is expressed in $rad/s$, and the input in $V$. The disturbance model is given by the exogenous system with system matrices:
\begin{align}
A_d = \begin{bmatrix}    0.9993 & \ 0.09994 \\
-0.0142 & \ 0.9985
\end{bmatrix}, \
B_d  = \begin{bmatrix}3.769 \cdot 10^{-5} \\  0.7535 \cdot 10^{-3} \end{bmatrix}, \
C_d = \begin{bmatrix}0 &  10^{5} \end{bmatrix}
\end{align}

Previous off-line experiences on the NCS framework lead to consider a maximum time delay $\tau_{max}$ slightly less than 200ms. Then, as commented in Section \ref{sec:intro} (Assumption \ref{as2}), the sensor period is chosen as $NT$=0.2s in order to ensure no packet disorder. Choosing $N=2$ allows the UGV to accurately track the path, as it will be shown in Section \ref{sec:SimExp}.

In this simulation, let us assume the packet dropout probability as $p_{sc}$=0.1 and $p_{ca}$=0.3 in \eqref{probability}.

Now, consider the continuous-time PID controller
$$u(t)=K_p \left[ e(t)+\frac{1}{T_i} \int_{0}^{t} e(\tau) d\tau \right] $$
where $K_p=6$ and $T_i=0.12$. The discretization of the above controller at period $T=0.1s$ and $NT=0.2s$  yields respectively
\begin{equation}\label{fast-rate}
G_r^T(z) =  \frac{6z-1}{z-1}
\end{equation}
\begin{equation}\label{slow-rate}
G_r^{NT}(z^N) = \frac{6z+4}{z-1}
\end{equation}

Also, the ZOH-discretized plant \eqref{eq:Gpppp} with sampling period $T$ leads to $G_p(z)= \frac{0.07082}{z - 0.445}$, which can be expressed in state-space model as \eqref{eq:model} with matrices:
\begin{align}
A_p = 0.4450, \quad B_p =  0.2500, \quad C_p = 0.2833
\end{align}
For robust performance analysis, time-varying model uncertainties $\Delta_{A,k}, \ \Delta_{B,k}$ on the form \eqref{eq:uncertt} are considered with $E_p=1$, $H_{p,A}=1$, $H_{p,B}=1$, where the scalar $\delta_{\Delta}$ will be later used as robust performance index.

The dual-rate controller with $N=2$ is obtained by means of (\ref{lento}) and (\ref{rapido}), leading to
\begin{eqnarray}\label{dual-rate1}
G_1^{NT}(z^N) & = & \frac{z^2-0.4734 z+0.05731}{z^2-1.191 z +0.1914} \\\label{dual-rate2}
G_2^T(z) & = & \frac{6.576 z^2-5.78 z+1.27}{z^2 -0.9578 z+0.2394}
\end{eqnarray}

The state-space matrices of the slow-rate and fast-rate controllers are respectively:
\begin{align}
&A_{\xi} = \begin{bmatrix}  1.1914  & \  -0.3829 \\
0.5000   & \       0
\end{bmatrix}, \ B_{\xi} = \begin{bmatrix} 1 \\ 0
\end{bmatrix}, \\ \nonumber
&\ C_{\xi} = \begin{bmatrix} 0.7180 & \  -0.2683 \end{bmatrix}, \ D_{\xi}=1, \\ \nonumber
& \\ \nonumber
&A_{\eta} = \begin{bmatrix}   0.9758  & \ -0.4788 \\
0.5000    & \       0
\end{bmatrix}, \ B_{\eta} = \begin{bmatrix} 1 \\ 0
\end{bmatrix}, \\ \nonumber
& C_{\eta} = \begin{bmatrix}  0.6367  & \ -0.6086 \end{bmatrix}, \ D_{\eta}=  6.5759
\end{align}

The parameter of the Kalman filter observed in simulation and experiments ranges between $K(\bar{N})=\begin{bmatrix} 14.1188 & \ 0 & \ 0.0001 \end{bmatrix}$ for $\bar{N}=N$ and $K(\bar{N})=\begin{bmatrix} 14.1202 & \ 0 & \ 0.0001 \end{bmatrix}$ for $\bar{N} \geq 6N$. For the robust performance analysis, we have taken the mean value $K(\bar{N})=\begin{bmatrix} 14.1195 & \ 0 & \ 0.0001 \end{bmatrix}$ since time variations in $K(\bar{N})$ can be considered negligible. 

\begin{figure}
	\centering
	\includegraphics[width=9cm, height=5cm]{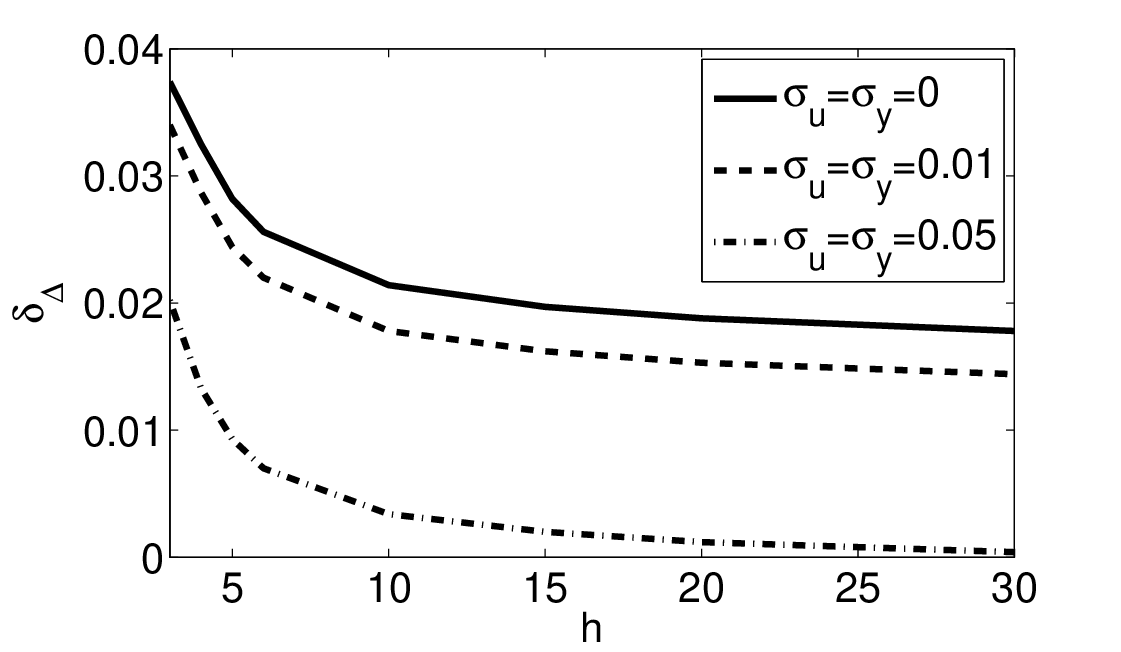}
	\caption{Robust index $\delta_{\Delta}$ with respect to the maximum number of consecutive measurement packet dropouts $h$}\label{fig:robust_vs_h}
\end{figure}

Fig. \ref{fig:robust_vs_h} depicts the closed-loop robustness margin $\delta_{\Delta}$ in \eqref{eq:uncertt}  obtained by solving the convex optimization problem described in Remark \ref{rem_opt} and Theorem \ref{teorema1}. The robust performance has been evaluated as a function of the maximum number of consecutive packet dropouts $h$ under three different choices of event-triggered parameters $\sigma_u$ and $\sigma_y$ with the same value for the sake of simplicity. On one hand, it is interesting to see that robustness decreases more slowly for larger values of $h$, which reveals that the designed NCS keeps a good robust performance even for large number of consecutive packet dropouts. On the other hand, it can be appreciated that the robustness decreases as long as the event-triggered parameters $\sigma_u, \sigma_y$ are higher, showing an existing trade-off between robust performance and bandwidth reduction. The event-triggered parameters $\Omega_u, \Omega_y$ in Theorem \ref{teorema1} have been chosen as $\Omega_u=\Omega_y=1$ since the input and output dimensions are equal to $1$, and therefore these parameters have no influence in robust performance by properly scaling $\sigma_u, \sigma_y$.


For the experimental setup, we have proposed the event-triggered parameters $\sigma_u = \sigma_y = 0$, $\Omega_u = \Omega_y = 1$, $\delta_y=0.01$ and $\delta_u=0.5$. The values of $\sigma_u, \sigma_y$ have been chosen to enhance the closed-loop robust performance following Remark \ref{rem_opt}. As a result, the parameter $\varepsilon=946.7456$ is obtained, which corresponds to a robustness margin $\delta_{\Delta}=0.0325$. The values of $\delta_u, \delta_y$ has been chosen to achieve some benefits in terms of reduction of bandwidth usage and energy consumption with a reasonable steady-state error, in light of the simulation and experimental results. The maximum number of consecutive packet dropouts will be set as $h=4$, and the reference to be followed will include a sequence of four right angles.

\subsection{Simulation results and experimental validation}\label{sec:SimExp}
In this section, the main advantages of the energy-efficient event-triggered control solution compared to different options for the time-triggered approach will be shown. The study will be focused on the
trade-off between resource usage and control properties.

\begin{figure}[H]
	\centering
	\includegraphics[width=7cm]{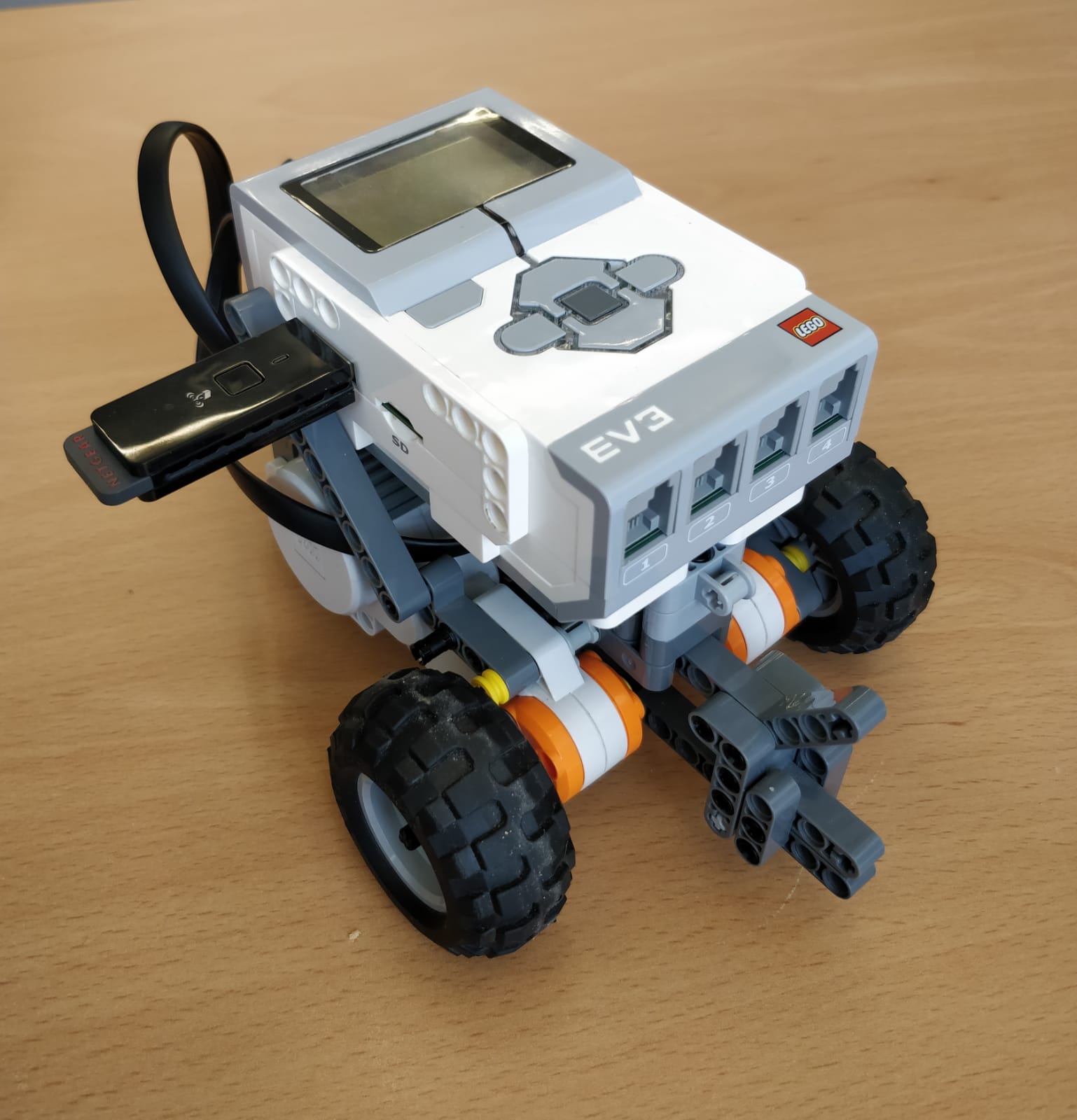}
	\caption{The Lego Mindstorms EV3 robot used in the implementation}
	\label{fig:robot}
\end{figure}

Simulation results will be experimentally validated by means of a test-bed platform, which is based on a two-wheel Lego Mindstorms EV3 robot equipped with a wifi-dongle to send and receive data-packages. A picture of the robot is shown in Figure \ref{fig:robot}. 

The proposed structure consists of two parts: i) at the remote side, a stationary, powerful computer is used to calculate all the set of control actions; ii) at the local side, the UGV, equipped with encoders and actuators, is employed. The control application is made in Matlab/Simulink\textregistered. The code programmed at the local side is kept as simple as possible because of the limited hardware capabilities of the UGV, and hence, it mainly includes User Datagram Protocol (UDP) send and receive blocks to wirelessly communicate with the remote side. The remote side includes these communication blocks as well, and the ones required to implement the control strategy. The model can be switched between experimental mode and simulation mode by changing the UDP blocks to a block representing the transfer function of each motor.

\begin{figure}[H]
\subfigure[Simulated result
]{\includegraphics[width = 2.7in]{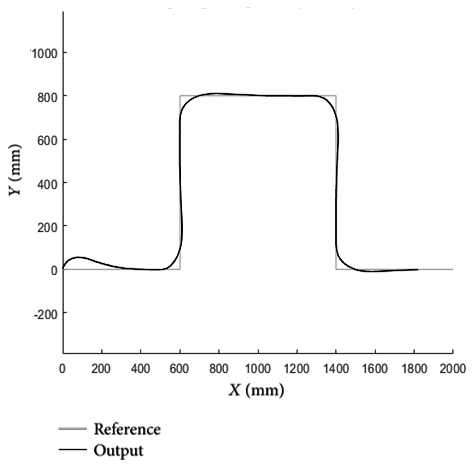}} 
\subfigure[Experimental result
]{\includegraphics[width = 2.7in]{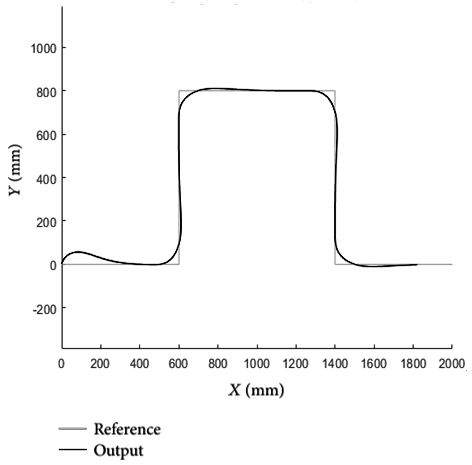}}
\caption{Comparison for the time-triggered single-rate \textit{T}=0.1s case (nominal performance; scenario \textit{b})}
\label{fig:01}
\end{figure}

\begin{figure}[H]
\subfigure[Simulated result
]{\includegraphics[width = 2.7in]{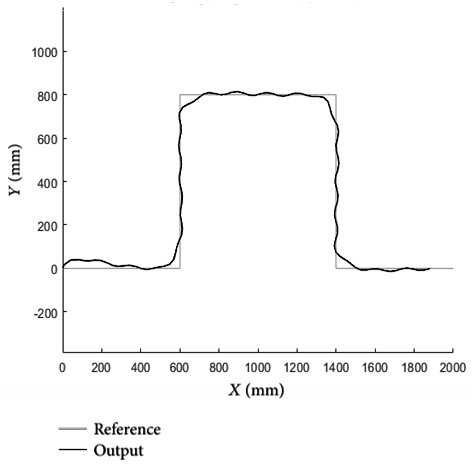}} 
\subfigure[Experimental result
]{\includegraphics[width = 2.7in]{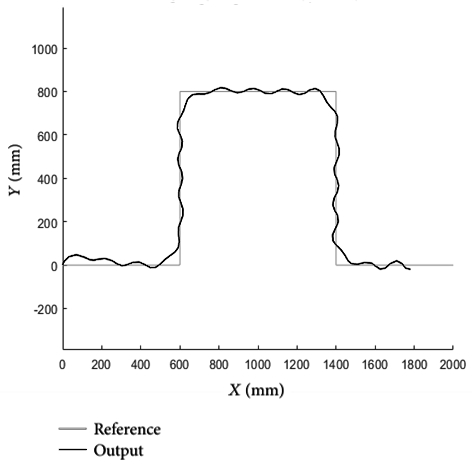}}
\caption{Comparison for the time-triggered single-rate \textit{NT}=0.2s case (scenario \textit{a})}
\label{fig:02}
\end{figure}

\begin{figure}[H]
\subfigure[Simulated result
]{\includegraphics[width = 2.7in]{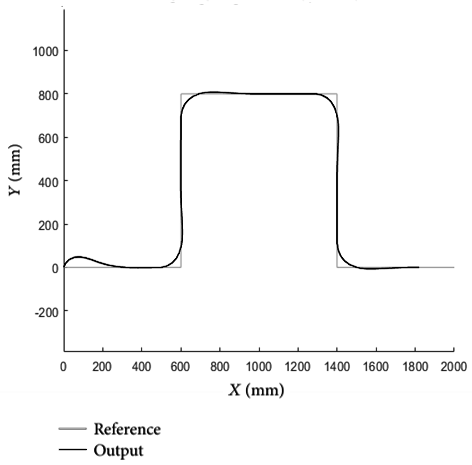}} 
\subfigure[Experimental result
]{\includegraphics[width = 2.7in]{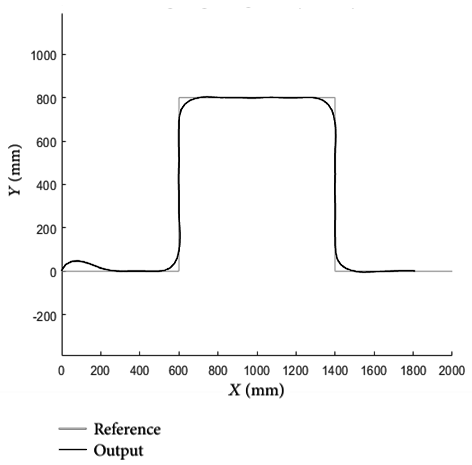}}
\caption{Comparison for the time-triggered dual-rate \textit{NT}=0.2s and \textit{T}=0.1s case (scenario \textit{c})}
\label{fig:dual}
\end{figure}

\begin{figure}[H]
\subfigure[Simulated result
]{\includegraphics[width = 2.7in]{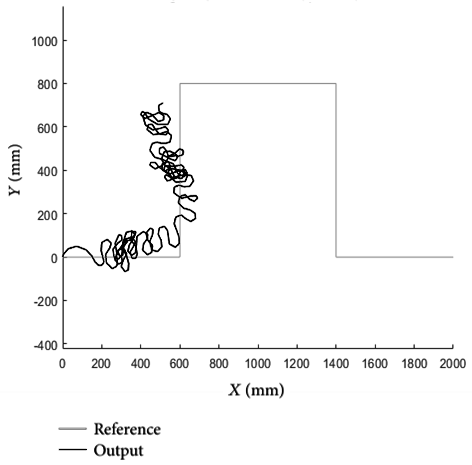}} 
\subfigure[Experimental result
]{\includegraphics[width = 2.7in]{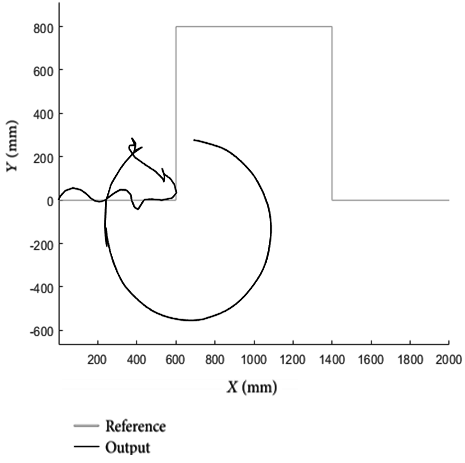}}
\caption{Comparison for the time-triggered dual-rate case, with delays and dropouts}
\label{fig:dual-delay-drops}
\end{figure}

\begin{figure}[H]
\subfigure[Simulated result
]{\includegraphics[width = 2.7in]{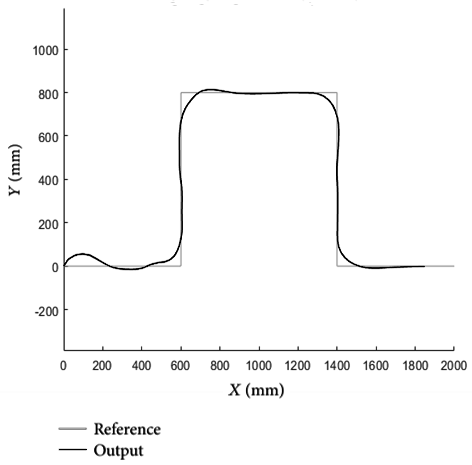}} 
\subfigure[Experimental result
]{\includegraphics[width = 2.7in]{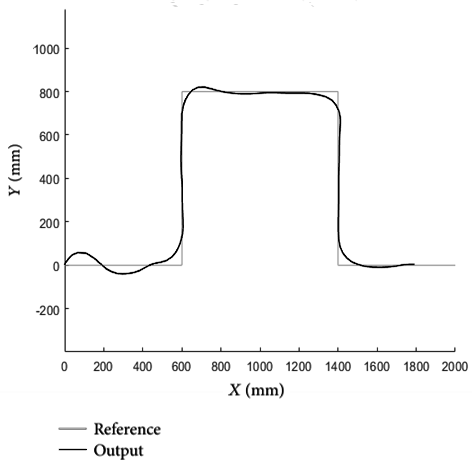}}
\caption{Comparison for the time-triggered dual-rate case, with TVDRKF and packet-based control, including delays, dropouts, noise and disturbances (scenario \textit{d})}
\label{fig:filter}
\end{figure}

\begin{figure}[H]
\subfigure[Simulated result
]{\includegraphics[width = 2.7in]{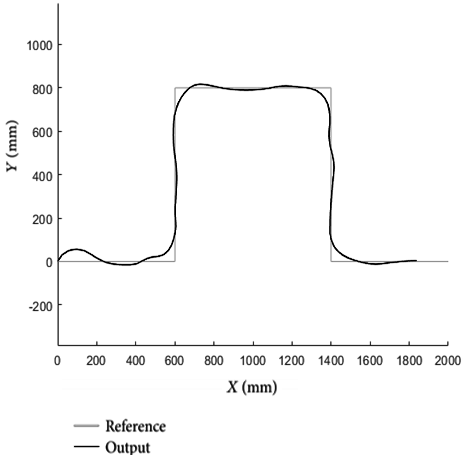}} 
\subfigure[Experimental result
]{\includegraphics[width = 2.7in]{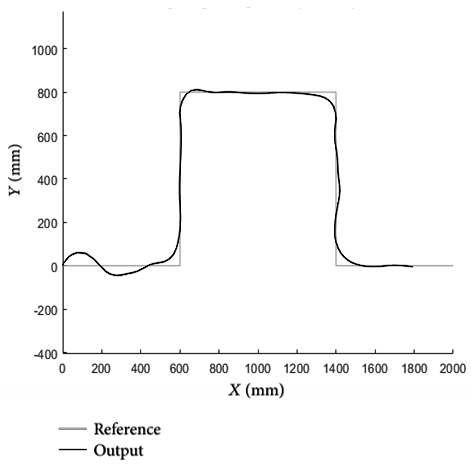}}
\caption{Comparison for the periodic event-triggered dual-rate case, with TVDRKF and packet-based control, including delays, dropouts, noise and disturbances (scenario \textit{e})}
\label{fig:event}
\end{figure}

The following results (Figures \ref{fig:01}-\ref{fig:event})  are obtained by running both working modes for different scenarios. In each figure, the simulated results are shown at the left hand, while the experimental validation is depicted at the right hand. 

Let us start the comparison considering a time-triggered single-rate control scenario with neither noise nor disturbance, and neither time-varying delays nor packet dropouts. Under these conditions, the UGV path tracking behavior achieved by the controller designed at period $NT$=0.2s in \eqref{slow-rate} (depicted in Figure \ref{fig:02}) is compared with the behavior obtained by the controller designed at period $T$=0.1s in \eqref{fast-rate} (shown in Figure \ref{fig:01}). The comparison illustrates a considerable performance worsening for the former case.

Let us consider the performance reached at period $T$ as the nominal, desired one. Taking into account the dual-rate controller in \eqref{dual-rate1}-\eqref{dual-rate2}, which injects control actions at period $T$ but measures system outputs at period $NT$, the time-triggered dual-rate control system is able to maintain a satisfactory control performance, very similar to the nominal one (as shown in Figure \ref{fig:dual}). Nevertheless, whether both time-varying delays and packet dropouts are included in the NCS, the performance is clearly worsened, becoming unstable (see in Figure \ref{fig:dual-delay-drops}).

Note also that the control system performance is clearly improved (as illustrated in Figure \ref{fig:filter}) by incorporating the TVDRKF into the time-triggered dual-rate control system (assuming delays and dropouts), together with the packet-based control. It can be appreciated that the obtained performance is very similar to the nominal one, despite the presence of measurement noise and external disturbances. 

Finally, event-triggered conditions are added to the NCS, and then, the system becomes a periodic event-triggered dual-rate control system. The performance obtained is similar to that reached by the nominal time-triggered version of the system (see in Figure \ref{fig:event}), but now a clear reduction of the number of transmitted packets is achieved, which leads to reducing resource usage (bandwidth, energy). This aspect will be analyzed in detail in Table \ref{tab:title}.


As shown in Figures \ref{fig:01}-\ref{fig:event}, the experimental results accurately validate the simulations. For this reason, the cost indexes presented in Section \ref{sec:Costfunc} will be directly evaluated from the experimentation. As the indexes may be negatively affected by the initialization part of the path, they will be calculated starting from iteration $k=20$. Table \ref{tab:title} summarizes the values obtained for the five different stable cases previously shown, that is:
\begin{itemize}
	\item \textit{a}: time-triggered single-rate control scenario at period $NT$ (in Figure \ref{fig:02}).
	\item \textit{b}: time-triggered single-rate control scenario at period $T$ (in Figure \ref{fig:01}).
	\item \textit{c}: time-triggered dual-rate control scenario (in Figure \ref{fig:dual}).
	\item \textit{d}: time-triggered dual-rate control scenario, adding TVDRKF, and packet-based control, including dropouts, delays, disturbances and noise (in Figure \ref{fig:filter}).
	\item \textit{e}: periodic event-triggered dual-rate control scenario, with TVDRKF, and packet-based control, including dropouts, delays, disturbances and noise (in Figure \ref{fig:event}).
\end{itemize}

\begin{table}[H]
	\centering
	\begin{tabular}{|c|ccccc|}
		\hline
		& \textit{a} & \textit{b} & \textit{c} &\textit{d} & \textit{e} \\ \hline
		$J_1$& 1328.3 & 1078.9 & 1085.0 & 1160.2 & 1189.9 \\
		$J_2$& 47.13 & 38.30 & 38.88 & 39.62 & 40.98 \\
		$J_3$& 22.4 & 22.0 & 22.0 & 22.0 & 22.0 \\
		$J_4$& 50\% & 100\% & 50\% & 50\% & 37.1\% \\ \hline
	\end{tabular}
	\caption {Cost indexes from the experimentation}
	\label{tab:title}
\end{table}

The underlying conclusions of Table \ref{tab:title} are:
\begin{itemize}
\item The desired nominal performance (scenario \textit{b}) presents the best (lowest) values for the cost indexes related to control performance, that is, $J_1, J_2, J_3$. The obtained values have been considered as the reference ones to carry out the comparison on control performance. 
\item Considerations for $J_1$: the single-rate case at period $NT$ (scenario \textit{a}) shows the worst $J_1$, since the desired trajectory is inaccurately followed by the UGV. The dual-rate case for scenario \textit{c} shows a similar $J_1$ than the nominal scenario \textit{b}. However, the dual-rate case for scenarios \textit{d} and \textit{e} worsen the index $J_1$ around 7\% and 10\%, respectively, compared to the nominal performance. 
\item Considerations for $J_2$: again, scenario \textit{a} (single-rate at $NT$) presents the worst value, being the dual-rate scenario \textit{c} very similar to the nominal scenario \textit{b}. The dual-rate scenarios \textit{d} and \textit{e} worsen the index $J_2$ around 3\% and 7\%, respectively, compared to the nominal performance. 
\item Considerations for $J_3$: this cost index presents the same value for the dual-rate cases \textit{c}, \textit{d}, and \textit{e} as for the nominal case \textit{b}, being 1\% worse for the single-rate scenario \textit{a} (at $NT$). 
\item Finally, to carry out the comparison on resource saving, $J_4$ will be analyzed. It presents the worst case in scenario \textit{b} (single-rate at $T$), being 50\% reduced by scenarios \textit{a}, \textit{c} and \textit{d} because of sampling the network twice slower (at $NT$, with $N=2$). The periodic event-triggered scenario \textit{e} reaches the best value for index $J_4$ with around 63\% of reduction, compared to the scenario \textit{b}.
\end{itemize}

As a summary, the periodic event-triggered dual-rate control scenario, with TVDRKF and packet-based control is able to significantly reduce resource usage (bandwidth, energy) with almost 63\% of reduction, while having a performance worsening of only up to 10\%, with regards to the nominal time-triggered case. This is in spite of existing dropouts, delays, disturbances and noise acting on the system. 

\section{Conclusions}\label{sec:conc}
This paper has addressed a formal analysis of the closed-loop stability and robust performance of an energy-efficient control in the framework of NCS, consisting in an event-triggered dual-rate control with a TVDRKF applied to a UGV. Different aspects, such as packet dropouts, time-varying model mismatches, disturbance rejection and event-triggered thresholds have been taken into account in the stability analysis. It has been illustrated that the robust performance is not significantly affected by the maximum number of consecutive packet dropouts in a wide range. In addition, the proposed control strategy has been experimentally validated by tracking the 2D position of a UGV in a wireless network. Despite existing typical network problems such as time-varying delays and packet dropouts, and considering disturbance and measurement noise, the control solution is able to achieve a satisfactory performance, while considerably reducing resource usage.

\section*{Acknowledgement}
This research was funded in part by grant by projects PGC2018-098719-B-I00 (MCIU/ AEI/ FEDER, UE), and RTI2018-096590-B-I00 (MCIU/ AEI/ FEDER, UE), and by European Commission as part of Project H2020-SEC-2016-2017 - Topic: SEC-20-BES-2016 - Id: 740736 - "C2 Advanced Multi-domain Environment and Live Observation Technologies" (CAMELOT). Part WP5 supported by Tekever ASDS, Thales Research and Technology, Viasat Antenna Systems, Universitat
Polit\`{e}cnica de Val\`{e}ncia, Funda\c{c}\~{a}o da Faculdade de Ci\^{e}ncias da Universidade de Lisboa, Minist\'{e}rio da Defensa Nacional - Marinha Portuguesa, Minist\'{e}rio da Administra\c{c}\~{a}o Interna Guarda Nacional Republicana.

\section*{References}

\bibliographystyle{abbrvnat}
\bibliography{bibliografia}

\begin{thebibliography}{50}
\providecommand{\natexlab}[1]{#1}
\providecommand{\url}[1]{\texttt{#1}}
\expandafter\ifx\csname urlstyle\endcsname\relax
  \providecommand{\doi}[1]{doi: #1}\else
  \providecommand{\doi}{doi: \begingroup \urlstyle{rm}\Url}\fi

\bibitem[Alcaina et~al.(2019)Alcaina, Cuenca, Salt, Zheng, and Tomizuka]{alcaina2019energy}
J.~Alcaina, {\'A}.~Cuenca, J.~Salt, M.~Zheng, and M.~Tomizuka.
\newblock Energy-efficient control for an unmanned ground vehicle in a wireless sensor network.
\newblock \emph{Journal of Sensors}, 2019:\penalty0 Article ID 7085915, 2019.

\bibitem[Andrade et~al.(2016)Andrade, Mansano, Godoy, and Porto]{andrade2016evaluation}
Y.~S. Andrade, R.~K. Mansano, E.~P. Godoy, and A.~J. Porto.
\newblock Evaluation of aperiodic control for energy saving in wireless networked control systems.
\newblock In \emph{2016 12th IEEE International Conference on Industry Applications (INDUSCON)}, pages 1--7. IEEE, 2016.

\bibitem[Bolot(1993)]{bolot1993end}
J.-C. Bolot.
\newblock End-to-end packet delay and loss behavior in the internet.
\newblock In \emph{ACM SIGCOMM Computer Communication Review}, volume~23, pages 289--298. ACM, 1993.

\bibitem[Boyd et~al.(1994)Boyd, El~Ghaoui, Feron, and Balakrishnan]{boyd1994linear}
S.~Boyd, L.~El~Ghaoui, E.~Feron, and V.~Balakrishnan.
\newblock \emph{Linear matrix inequalities in system and control theory}, volume~15.
\newblock SIAM, 1994.

\bibitem[Cervin et~al.(2003)Cervin, Henriksson, Lincoln, Eker, and Arzen]{cervin2003does}
A.~Cervin, D.~Henriksson, B.~Lincoln, J.~Eker, and K.-E. Arzen.
\newblock {How does control timing affect performance? Analysis and simulation of timing using Jitterbug and TrueTime}.
\newblock \emph{IEEE Control Systems}, 23\penalty0 (3):\penalty0 16--30, 2003.

\bibitem[Cloosterman et~al.(2009)Cloosterman, Van~de Wouw, Heemels, and Nijmeijer]{cloosterman2009stability}
M.~B. Cloosterman, N.~Van~de Wouw, W.~Heemels, and H.~Nijmeijer.
\newblock Stability of networked control systems with uncertain time-varying delays.
\newblock \emph{IEEE Transactions on Automatic Control}, 54\penalty0 (7):\penalty0 1575--1580, 2009.

\bibitem[Cuenca et~al.(2018{\natexlab{a}})Cuenca, Alcaina, Salt, Casanova, and Piz{\'a}]{cuenca2018packet}
A.~Cuenca, J.~Alcaina, J.~Salt, V.~Casanova, and R.~Piz{\'a}.
\newblock {A packet-based dual-rate PID control strategy for a slow-rate sensing Networked Control System}.
\newblock \emph{ISA Transactions}, 76:\penalty0 155--166, 2018{\natexlab{a}}.

\bibitem[Cuenca et~al.(2018{\natexlab{b}})Cuenca, Antunes, Castillo, Garc{\'\i}a, Khashooei, and Heemels]{cuenca2018periodic}
A.~Cuenca, D.~J. Antunes, A.~Castillo, P.~Garc{\'\i}a, B.~A. Khashooei, and W.~Heemels.
\newblock Periodic event-triggered sampling and dual-rate control for a wireless networked control system with applications to $\text{UAV}$s.
\newblock \emph{IEEE Transactions on Industrial Electronics}, 66\penalty0 (4):\penalty0 3157--3166, 2018{\natexlab{b}}.

\bibitem[Cuenca et~al.(2018{\natexlab{c}})Cuenca, Zheng, Tomizuka, and S{\'a}nchez]{cuenca2018non}
{\'A}.~Cuenca, M.~Zheng, M.~Tomizuka, and S.~S{\'a}nchez.
\newblock Non-uniform multi-rate estimator based periodic event-triggered control for resource saving.
\newblock \emph{Information Sciences}, 459:\penalty0 86--102, 2018{\natexlab{c}}.

\bibitem[Ding et~al.(2020)Ding, Xie, and Liu]{ding2020event}
S.~Ding, X.~Xie, and Y.~Liu.
\newblock Event-triggered static/dynamic feedback control for discrete-time linear systems.
\newblock \emph{Information Sciences}, 524:\penalty0 33--45, 2020.

\bibitem[Dolk et~al.(2017)Dolk, Ploeg, and Heemels]{dolk2017event}
V.~S. Dolk, J.~Ploeg, and W.~M.~H. Heemels.
\newblock Event-triggered control for string-stable vehicle platooning.
\newblock \emph{IEEE Transactions on Intelligent Transportation Systems}, 18\penalty0 (12):\penalty0 3486--3500, 2017.

\bibitem[Dolz et~al.(2016)Dolz, Pe{\~n}arrocha, and Sanchis]{dolz2016networked}
D.~Dolz, I.~Pe{\~n}arrocha, and R.~Sanchis.
\newblock Networked gain-scheduled fault diagnosis under control input dropouts without data delivery acknowledgment.
\newblock \emph{International Journal of Robust and Nonlinear Control}, 26\penalty0 (4):\penalty0 737--758, 2016.

\bibitem[Gonz{\'a}lez(2021)]{gonzalez2021weighted}
A.~Gonz{\'a}lez.
\newblock A weighted distributed predictor-feedback control synthesis for interconnected time delay systems.
\newblock \emph{Information Sciences}, 543:\penalty0 367--381, 2021.

\bibitem[Gonzalez et~al.(2019)Gonzalez, Balaguer, Garcia, and Cuenca]{gonzalez2019gain}
A.~Gonzalez, V.~Balaguer, P.~Garcia, and A.~Cuenca.
\newblock Gain-scheduled predictive extended state observer for time-varying delays systems with mismatched disturbances.
\newblock \emph{ISA Transactions}, 84:\penalty0 206--213, 2019.

\bibitem[Gonz{\'a}lez et~al.(2019)Gonz{\'a}lez, Cuenca, Balaguer, and Garc{\'\i}a]{gonzalez2019event}
A.~Gonz{\'a}lez, A.~Cuenca, V.~Balaguer, and P.~Garc{\'\i}a.
\newblock Event-triggered predictor-based control with gain-scheduling and extended state observer for networked control systems.
\newblock \emph{Information Sciences}, 491:\penalty0 90--108, 2019.

\bibitem[Guo and Chen(2005)]{guo2005disturbance}
L.~Guo and W.-H. Chen.
\newblock Disturbance attenuation and rejection for systems with nonlinearity via dobc approach.
\newblock \emph{International Journal of Robust and Nonlinear Control: IFAC-Affiliated Journal}, 15\penalty0 (3):\penalty0 109--125, 2005.

\bibitem[Guo et~al.(2020)Guo, Yu, and Hao]{guo2020event}
L.~Guo, H.~Yu, and F.~Hao.
\newblock Event-triggered control for stochastic networked control systems against denial-of-service attacks.
\newblock \emph{Information Sciences}, 527:\penalty0 51--69, 2020.

\bibitem[Han and Gu(2001)]{han2001robust}
Q.-L. Han and K.~Gu.
\newblock On robust stability of time-delay systems with norm-bounded uncertainty.
\newblock \emph{IEEE Transactions on Automatic Control}, 46\penalty0 (9):\penalty0 1426--1431, 2001.

\bibitem[Han et~al.(2018)Han, Xu, Chen, Huang, and Zhao]{han2018energy}
Z.~Han, N.~Xu, H.~Chen, Y.~Huang, and B.~Zhao.
\newblock Energy-efficient control of electric vehicles based on linear quadratic regulator and phase plane analysis.
\newblock \emph{Applied Energy}, 213:\penalty0 639--657, 2018.

\bibitem[Hao et~al.(2019)Hao, Liu, and Zhou]{hao2019output}
S.~Hao, T.~Liu, and B.~Zhou.
\newblock Output feedback anti-disturbance control of input-delayed systems with time-varying uncertainties.
\newblock \emph{Automatica}, 104:\penalty0 8--16, 2019.

\bibitem[Heemels et~al.(2012)Heemels, Donkers, and Teel]{heemels2012periodic}
W.~H. Heemels, M.~Donkers, and A.~R. Teel.
\newblock Periodic event-triggered control for linear systems.
\newblock \emph{IEEE Transactions on Automatic Control}, 58\penalty0 (4):\penalty0 847--861, 2012.

\bibitem[Khashooei et~al.(2017)Khashooei, van Eekelen, Antunes, and Heemels]{khashooei2017suboptimal}
B.~A. Khashooei, B.~van Eekelen, D.~J. Antunes, and W.~M.~H. Heemels.
\newblock Suboptimal event-triggered control over unreliable communication links with experimental validation.
\newblock In \emph{2017 3rd International Conference on Event-Based Control, Communication and Signal Processing (EBCCSP)}, pages 1--6. IEEE, 2017.

\bibitem[K{\"u}hne et~al.(2018)K{\"u}hne, P{\"o}schke, and Schulte]{kuhne2018fault}
P.~K{\"u}hne, F.~P{\"o}schke, and H.~Schulte.
\newblock Fault estimation and fault-tolerant control of the $\text{FAST NREL 5-MW}$ reference wind turbine using a proportional multi-integral observer.
\newblock \emph{International Journal of Adaptive Control and Signal Processing}, 32\penalty0 (4):\penalty0 568--585, 2018.

\bibitem[Li et~al.(2002)Li, Shah, and Chen]{li2002analysis}
D.~Li, S.~L. Shah, and T.~Chen.
\newblock Analysis of dual-rate inferential control systems.
\newblock \emph{Automatica}, 38\penalty0 (6):\penalty0 1053--1059, 2002.

\bibitem[Li and Shi(2014)]{li2014network}
H.~Li and Y.~Shi.
\newblock Network-based predictive control for constrained nonlinear systems with two-channel packet dropouts.
\newblock \emph{IEEE Transactions on Industrial Electronics}, 61\penalty0 (3):\penalty0 1574--1582, 2014.

\bibitem[Li et~al.(2020)Li, Wang, Zhai, and Fei]{li2020event}
T.~Li, T.~Wang, J.~Zhai, and S.~Fei.
\newblock Event-triggered observer-based robust $\text{H}_{\infty}$ control for networked control systems with unknown disturbance.
\newblock \emph{International Journal of Robust and Nonlinear Control}, 30\penalty0 (7):\penalty0 2671--2688, 2020.

\bibitem[Liu et~al.(2015)Liu, Zhang, Yu, Liu, and Chen]{liu2015new}
A.~Liu, W.~A. Zhang, L.~Yu, S.~Liu, and M.~Z. Chen.
\newblock New results on stabilization of networked control systems with packet disordering.
\newblock \emph{Automatica}, 52:\penalty0 255--259, 2015.

\bibitem[Liu et~al.(2017)Liu, Zhang, Chen, and Yu]{liu2017networked}
A.~Liu, W.~A. Zhang, B.~Chen, and L.~Yu.
\newblock Networked filtering with markov transmission delays and packet disordering.
\newblock \emph{IET Control Theory \& Applications}, 12\penalty0 (5):\penalty0 687--693, 2017.

\bibitem[Lundgren(2003)]{lundgren2003path}
M.~Lundgren.
\newblock Path tracking and obstacle avoidance for a miniature robot.
\newblock \emph{Ume{\aa} University, Ume{\aa}, Master Thesis}, 2003.

\bibitem[Luo et~al.(2021)Luo, Xiao, Cao, Li, and Lin]{luo2021event}
Y.~Luo, X.~Xiao, J.~Cao, A.~Li, and G.~Lin.
\newblock Event-triggered guaranteed cost consensus control for second-order multi-agent systems based on observers.
\newblock \emph{Information Sciences}, 546:\penalty0 283--297, 2021.

\bibitem[Park et~al.(2019)Park, Nah, Choi, Yoon, and Park]{park2019robust}
B.~Park, J.~Nah, J.-Y. Choi, I.-J. Yoon, and P.~Park.
\newblock Robust wireless sensor and actuator networks for networked control systems.
\newblock \emph{Sensors}, 19\penalty0 (7):\penalty0 1535, 2019.

\bibitem[Pe{\~n}arrocha et~al.(2012)Pe{\~n}arrocha, Sanchis, and Romero]{penarrocha2012state}
I.~Pe{\~n}arrocha, R.~Sanchis, and J.~A. Romero.
\newblock State estimator for multisensor systems with irregular sampling and time-varying delays.
\newblock \emph{International Journal of Systems Science}, 43\penalty0 (8):\penalty0 1441--1453, 2012.

\bibitem[Raghunathan et~al.(2002)Raghunathan, Schurgers, Park, and Srivastava]{raghunathan2002energy}
V.~Raghunathan, C.~Schurgers, S.~Park, and M.~B. Srivastava.
\newblock Energy-aware wireless microsensor networks.
\newblock \emph{IEEE Signal Processing Magazine}, 19\penalty0 (2):\penalty0 40--50, 2002.

\bibitem[Sala et~al.(2009)Sala, Cuenca, and Salt]{sala2009retunable}
A.~Sala, A.~Cuenca, and J.~Salt.
\newblock A retunable $\text{PID}$ multi-rate controller for a networked control system.
\newblock \emph{Information Sciences}, 179\penalty0 (14):\penalty0 2390--2402, 2009.

\bibitem[Salt and Albertos(2005)]{salt2005mbm}
J.~Salt and P.~Albertos.
\newblock {Model-Based Multirate Controllers Design}.
\newblock \emph{IEEE Transactions on Control Systems Technologies}, 13\penalty0 (6):\penalty0 988--997, Nov. 2005.

\bibitem[Shah et~al.(2020)Shah, Mehta, Patel, and Bartoszewicz]{shah2020event}
D.~Shah, A.~Mehta, K.~Patel, and A.~Bartoszewicz.
\newblock Event-triggered discrete higher-order smc for networked control system having network irregularities.
\newblock \emph{IEEE Transactions on Industrial Informatics}, 16\penalty0 (11):\penalty0 6837--6847, 2020.

\bibitem[Simkoff et~al.(2020)Simkoff, Lejarza, Kelley, Tsay, and Baldea]{simkoff2020process}
J.~M. Simkoff, F.~Lejarza, M.~T. Kelley, C.~Tsay, and M.~Baldea.
\newblock Process control and energy efficiency.
\newblock \emph{Annual Review of Chemical and Biomolecular Engineering}, 11:\penalty0 423--445, 2020.

\bibitem[Simon(2006)]{simon2006optimal}
D.~Simon.
\newblock \emph{Optimal state estimation: Kalman, $H_{\infty}$, and nonlinear approaches}.
\newblock John Wiley \& Sons, 2006.

\bibitem[{\v{S}}irok{\`y} et~al.(2011){\v{S}}irok{\`y}, Oldewurtel, Cigler, and Pr{\'\i}vara]{vsiroky2011experimental}
J.~{\v{S}}irok{\`y}, F.~Oldewurtel, J.~Cigler, and S.~Pr{\'\i}vara.
\newblock Experimental analysis of model predictive control for an energy efficient building heating system.
\newblock \emph{Applied Energy}, 88\penalty0 (9):\penalty0 3079--3087, 2011.

\bibitem[Suh(2008)]{suh2008stability}
Y.~S. Suh.
\newblock Stability and stabilization of nonuniform sampling systems.
\newblock \emph{Automatica}, 44\penalty0 (12):\penalty0 3222--3226, 2008.

\bibitem[Trimpe and D'Andrea(2014)]{trimpe2014event}
S.~Trimpe and R.~D'Andrea.
\newblock Event-based state estimation with variance-based triggering.
\newblock \emph{IEEE Transactions on Automatic Control}, 59\penalty0 (12):\penalty0 3266--3281, 2014.

\bibitem[Tripathy et~al.(2019)Tripathy, Chamanbaz, and Bouffanais]{tripathy2019robust}
N.~S. Tripathy, M.~Chamanbaz, and R.~Bouffanais.
\newblock Robust stabilization of resource limited networked control systems under denial-of-service attack.
\newblock In \emph{2019 IEEE 58th Conference on Decision and Control (CDC)}, pages 7683--7689. IEEE, 2019.

\bibitem[Tripathy et~al.(2020)Tripathy, Kar, Chamanbaz, and Bouffanais]{tripathy2020robust}
N.~S. Tripathy, I.~N. Kar, M.~Chamanbaz, and R.~Bouffanais.
\newblock Robust stabilization of a class of nonlinear systems via aperiodic sensing and actuation.
\newblock \emph{IEEE Access}, 8:\penalty0 157403--157417, 2020.

\bibitem[Xia and Zhang(2010)]{xia2010energy}
X.~Xia and J.~Zhang.
\newblock {Energy efficiency and control systems--from a POET perspective}.
\newblock \emph{IFAC Proceedings Volumes}, 43\penalty0 (1):\penalty0 255--260, 2010.

\bibitem[Xia and Zhang(2016)]{xia2016industrial}
X.~Xia and L.~Zhang.
\newblock Industrial energy systems in view of energy efficiency and operation control.
\newblock \emph{Annual Reviews in Control}, 42:\penalty0 299--308, 2016.

\bibitem[Zhang et~al.(2021)Zhang, Zhang, Sun, and Gao]{zhang2021leader}
J.~Zhang, H.~Zhang, S.~Sun, and Z.~Gao.
\newblock Leader-follower consensus control for linear multi-agent systems by fully distributed edge-event-triggered adaptive strategies.
\newblock \emph{Information Sciences}, 555:\penalty0 314--338, 2021.

\bibitem[Zhang et~al.(2019)Zhang, Han, Ge, Ding, Ding, Yue, and Peng]{zhang2019networked}
X.-M. Zhang, Q.-L. Han, X.~Ge, D.~Ding, L.~Ding, D.~Yue, and C.~Peng.
\newblock Networked control systems: a survey of trends and techniques.
\newblock \emph{IEEE/CAA Journal of Automatica Sinica}, 7\penalty0 (1):\penalty0 1--17, 2019.

\bibitem[Zheng et~al.(2016{\natexlab{a}})Zheng, Chen, and Tomizuka]{zheng2016extended}
M.~Zheng, X.~Chen, and M.~Tomizuka.
\newblock Extended state observer with phase compensation to estimate and suppress high-frequency disturbances.
\newblock In \emph{American Control Conf. (ACC)}, pages 3521--3526. IEEE, 2016{\natexlab{a}}.

\bibitem[Zheng et~al.(2016{\natexlab{b}})Zheng, Sun, and Tomizuka]{zheng2016multi}
M.~Zheng, L.~Sun, and M.~Tomizuka.
\newblock {Multi-rate observer based sliding mode control with frequency shaping for vibration suppression beyond Nyquist frequency}.
\newblock \emph{IFAC-PapersOnLine}, 49\penalty0 (21):\penalty0 13--18, 2016{\natexlab{b}}.

\bibitem[Zou et~al.(2016)Zou, Lin, Feng, and Chen]{zou2016energy}
T.~Zou, S.~Lin, Q.~Feng, and Y.~Chen.
\newblock Energy-efficient control with harvesting predictions for solar-powered wireless sensor networks.
\newblock \emph{Sensors}, 16\penalty0 (1):\penalty0 53, 2016.

\end{thebibliography}

\end{document}